\documentstyle[epsfig]{mn}

\def\gs{\mathrel{\raise0.35ex\hbox{$\scriptstyle >$}\kern-0.6em
\lower0.40ex\hbox{{$\scriptstyle \sim$}}}}
\def\ls{\mathrel{\raise0.35ex\hbox{$\scriptstyle <$}\kern-0.6em
\lower0.40ex\hbox{{$\scriptstyle \sim$}}}}

\begin{document}

\title
[Star-formation history] 
{The history of star formation in dusty galaxies}
\author
[A.\,W. Blain, Ian Smail, R.\,J. Ivison \& J.-P. Kneib]
{
A.\,W. Blain,$^1$ Ian Smail,$^{\! 2}$ R.\,J.~Ivison$^{3,4}$ and J.-P. Kneib$^5$\\
$^1$ Cavendish Laboratory, Madingley Road, Cambridge, CB3 OHE, UK.\\
$^2$ Department of Physics, University of Durham, South Road, Durham,
DH1 3LE, UK.\\
$^3$ Institute for Astronomy, Department of Physics \& Astronomy, University
of Edinburgh, Blackford Hill, Edinburgh, EH9 3HJ, UK.\\
$^4$ Department of Physics \& Astronomy, University College London, Gower 
Street, London, WC1E 6BT, UK.\\
$^5$ Observatoire Midi-Pyr\'en\'ees, 14 Avenue E. Belin, 31400 Toulouse, France.
}
\maketitle

\begin{abstract}
A population of distant dusty galaxies emitting in the submillimetre waveband 
has recently been detected using the Submillimetre Common-User Bolometer 
Array (SCUBA) camera on the James Clerk Maxwell Telescope (JCMT). This 
population can be used to trace the amount of high-redshift star-formation 
activity that is obscured from view in the optical waveband by dust, and so is 
missing from existing inventories of star formation in the distant Universe. By
including this population we can construct a complete and consistent picture 
of the history of star formation. The evolution of obscured star formation at 
redshifts less than unity is constrained by mid- and far-infrared counts of dusty 
galaxies.  Activity increases with redshift $z$ as $(1+z)^\gamma$ with 
$\gamma \sim 4$, consistent with the form of evolution found in the optical 
waveband by the Canada--France Redshift Survey (CFRS) to $z\ls 1$. The form 
of evolution at higher redshifts is constrained by both faint SCUBA counts 
and the intensity of background radiation in the 
millimetre/submillimetre waveband. We find that the total amount of energy 
emitted by dusty galaxies is about four times greater than that inferred from 
restframe ultraviolet observations, and that a larger fraction of this energy is 
emitted at high redshifts. The simplest explanation for these results is that a 
large population of luminous strongly-obscured sources at redshifts $z\ls 5$ is 
missing from optical surveys. We discuss the possible contribution of obscured 
active galactic nuclei to the submillimetre-wave background and counts. More 
accurate constraints on the history of star formation will be provided by 
determinations of the counts in several submillimetre wavebands and crucially
by a reliable redshift distribution of the detected galaxies.
\end{abstract}  


\section{Introduction}

A history of star formation in galaxies over roughly the last half of
the Hubble time has been inferred with reasonable 
accuracy by combining the star-formation rate (SFR) in the local universe 
(Gallego et al.\ 1996; Gronwall 1998; Tresse \& Maddox 1997; Treyer et al.\ 1998), 
the results of spectroscopic redshift 
surveys out to $z\sim 1$ (Lilly et al.\ 1996) and the results of deep photometric 
surveys, such as the {\it Hubble Deep Field}, in which Lyman-dropout selection 
techniques (Steidel et al.\ 1996a) are used to determine photometric redshifts: 
see Madau et al.\ (1996) and Madau, Della Valle \& Panagia (1998). 
This history of unobscured star formation in galaxies can be described 
adequately by a power-law increase in the comoving density of star-formation 
activity with redshift $z$ of the form $(1+z)^\gamma$ with $\gamma\sim 4$ out 
to $z\simeq 1$, followed by a decrease at higher redshifts. Observations made 
in the near-infrared waveband by Connolly et al.\ (1997) have provided further 
support for the suggestion that the SFR peaks at $z\sim 1$--2. This standard 
picture of the star-formation history is shown in Fig.\,1. It accords with the form 
of evolution of the luminosity density inferred from observations of radio 
sources (Cram 1998; Dunlop 1998) and quasars (Boyle \& Terlevich 1998). 

This work has greatly increased our knowledge of the evolution of star formation 
activity in the Universe; however, the points plotted in Fig.\,1 in fact only 
represent firm lower limits to the true SFR at each epoch. Extinction due to any 
interstellar dust in the optically selected observed galaxies suppresses their 
flux densities in the rest-frame optical and ultraviolet wavebands, and so leads 
to an underestimation of their SFR, sometimes by quite considerable factors
($\gs 10$). 
Using estimates of obscuration in Lyman-dropout galaxies derived from 
restframe ultraviolet spectra, Pettini et al.\ (1998a,b) have recently revised 
the global SFR at redshifts greater than 2 upwards by a factor of 
about 5, as shown by the open boxes in Fig.\,1. However, it is difficult to 
correct optical and ultraviolet spectra reliably for the effects of dust, and so 
these revisions are rather uncertain. In contrast, the associated rest-frame 
mid- and far-infrared thermal emission from the interstellar dust grains that are 
heated by absorbing optical and ultraviolet photons can be detected directly 
and used to place limits on the dust mass and SFR of the galaxies. 

The first observations of dust emission from external galaxies were made by the 
{\it IRAS} satellite. Nearby 
galaxies selected on the grounds of blue colours in the optical waveband were 
found to emit typically 60 per cent of their total luminosity in the far-infrared 
waveband (Mazzarella \& Balzano 1986). The redshift distribution of the 
population of galaxies detected at a wavelength of 60\,$\mu$m by {\it IRAS} 
extends out to $z\simeq 0.2$, and is consistent with pure luminosity 
evolution of the form $(1+z)^{3.1}$ (Saunders et al.\ 1990; Oliver, 
Rowan-Robinson \& Saunders 1992). More recent observations at wavelengths 
of 15\,$\mu$m and 175\,$\mu$m using the more sensitive {\it ISO} satellite 
(Kawara et al.\ 1997; Rowan-Robinson et al.\ 1997; Hammer \& Flores 1998; 
Lagache et al.\ 1998) 
have confirmed that strong evolution continues out to $z\sim 1$, a result that 
is broadly consistent with the form of evolution derived from optical observations 
(Lilly et al.\ 1996).   However, although they trace the star-formation
history over at least half the Hubble time, these observations do not
strongly constrain the SFR in the early Universe, an era crucial for
distinguishing between competing galaxy formation models (Cole et al.\ 1994).
One of the most promising routes to probe this regime are deep observations at
still longer wavelengths, in the millimetre and submillimetre wavebands.

Early bounds on the intensity of background radiation in the submillimetre and 
far-infrared wavebands from the FIRAS instrument on the {\it COBE} satellite
(Wright et al.\ 1994; Fixsen et al.\ 1996) 
provided limits on the amount and epoch of obscured star 
formation activity in an hierarchical model (Blain \& Longair 1993b). Much more 
information about the role of dust was provided by Puget et al.\ (1996), who 
detected a considerably larger isotropic non-Galactic background radiation 
signal in the same data by using a different model of Galactic emission. 
The results indicate that a large amount of far-infrared emission must be 
produced by dust at $z\gs 1$ (Burigana et al.\ 1997). 
Recently, this early estimate of the background radiation intensity has been in 
large part confirmed by 
independent analyses (Guiderdoni et al.\ 1997; Fixsen et al.\ 1998), 
and by complimentary data obtained 
using the DIRBE instrument (Hauser et al.\ 1998; 
Schlegel, Finkbeiner \& Davis 1998). 
A significant amount of star formation activity at redshifts greater
than unity, at least in dense environments, is apparently required to explain 
observations of metal enrichment in intracluster gas at $z\sim0.5$ (Mushotzky \& 
Loewenstein 1997). 

The population of submillimetre-luminous galaxies that produce the
far-infrared/submillimetre-wave background signal were detected by Smail, 
Ivison \& Blain (1997, hereafter SIB). 
The counts of faint galaxies in the submillimetre 
waveband are very sensitive to the form of galaxy evolution at high redshift
because the $k$--corrections expected for distant galaxies are very large and 
negative. Hence high-redshift galaxies are likely to make a substantial 
contribution to the counts (Blain \& Longair 1993a), which in turn provide a 
sensitive test of galaxy evolution models in the early Universe.

\begin{figure}
\begin{center}
\epsfig{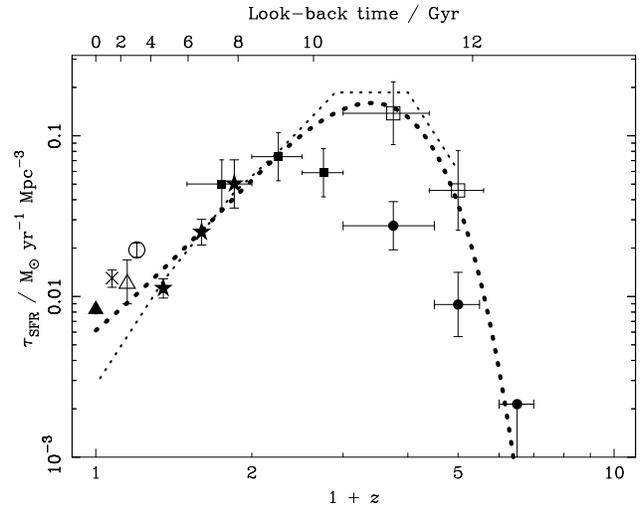} \hskip 7mm
\end{center}
\caption{The star-formation rate (SFR) as a function of redshift 
$\tau_{\rm SFR}(z)$, as inferred from ultraviolet/optical/near-infrared 
observations by Gallego et al.\ (1996; filled triangle), Gronwall (1998; diagonal 
cross), Treyer et al.\ (1998; open triangle), Tresse \& Maddox (1998; empty circle), 
Lilly et al.\ (1996; stars), Connolly et al. (1997; filled squares) and Madau et al.\ 
(1996; filled circles). A modified high-redshift SFR, inferred after a correction for 
dust extinction by Pettini et al.\ (1998a), is shown by the empty squares. The thick 
dotted curve is the SFR inferred from radio data by Dunlop (1998). The evolution 
of the luminosity density of AGN at a wavelength of 280\,nm (Boyle \& Terlevich 
1998) is shown by the thin dotted line; it is converted into an equivalent SFR, 
$\tau_{\rm SFR}$ assuming that 
$\tau_{\rm SFR}=10^{-2}$\,M$_\odot$\,yr$^{-1}$\,Mpc$^{-3}$ is equivalent to a 
280-nm luminosity density of $1.7 \times 10^{17}$\,W\,Hz$^{-1}$\,Mpc$^{-3}$. 
}
\end{figure}

SIB exploited the especially strong magnification bias expected in rich clusters 
of galaxies in the submillimetre waveband (Blain 1997) to increase the 
detectability of the population of background galaxies. Unless the cluster 
galaxies have extremely high and undetected SFRs, the contribution of
flux densities from galaxies within the lensing clusters should be at least an 
order of magnitude less than that from background sources. The 
observations were carried out using the sensitive SCUBA 
camera at the James Clerk Maxwell Telescope  
(Holland et al.\ 1999).
The results of the completed cluster survey and its
follow up are described elsewhere (Ivison et al.\ 1998; Smail et al.\ 1998;
Blain et al.\ submitted; Smail et al. in preparation). 
The first results of other blank-field surveys (Barger et al.\ 
1998; Eales et al.\ 1998; Holland et al.\ 1998b; Hughes et al.\ 1998) are 
now available. 

We use these 850-$\mu$m observations and other sub-millimetre/far-infrared 
observations at wavelengths of 175\,$\mu$m (Kawara et al.\ 1997) and 
2.8\,mm (Wilner \& Wright 1997), the intensity of 
background radiation in the millimetre/submillimetre/far-infrared wavebands, and 
existing information about the evolution of dusty galaxies at low redshifts 
derived from 60-$\mu$m observations (Oliver et al.\ 1992) to investigate the 
history of star-formation activity in the Universe that is obscured from view in 
the optical waveband by dust. Throughout the paper we use the comoving 
density of star-formation activity to represent the luminosity density of the 
Universe, in order that the results of observations in the optical and 
submillimetre wavebands can be readily compared. However, non-stellar 
emission from active galactic nuclei (AGN) contribute to the total emitted 
radiation in the submillimetre waveband. This issue is discussed in more detail 
in Section\,5.

We review the relevant observations and existing limits in Section\,2, outline our 
strategy for determining the star formation history in Section\,3, and discuss 
the self-consistency of the results in Section\,4. In Section\,5 we present the
resulting star formation histories and review them in the context of optical 
observations. In Section\,6 we investigate the most promising routes to 
obtaining better constraints on the history of star-formation activity in the 
millimetre/submillimetre and far-infrared wavebands: see also Blain (1998c).  
A value of Hubble's constant 
$H_0=50$\,km\,s$^{-1}$\,Mpc$^{-1}$, a density parameter $\Omega_0=1$ and a 
cosmological constant $\Omega_\Lambda=0$ are assumed. 

\section{Observations}

\subsection{Mid/far-infrared source counts}

The whole sky was observed at a wavelength of 60\,$\mu$m in the {\it IRAS} 
survey, and counts of galaxies were determined to various flux density limits 
(Hacking \& Houck 1987; Rowan-Robinson et al.\ 1990; Saunders et al.\ 1990; 
Gregorich et al.\ 1995; Bertin, Dennefeld \& Moshir 1997). 
At this wavelength the emission from galaxies is 
dominated by the thermal radiation of dust grains, heated by absorbing the 
optical and ultraviolet light from both young high-mass stars and AGN. The
most luminous class of sources detected by {\it IRAS} appear to be powered 
by stars and AGN in the ratio 2:1 (Sanders \& Mirabel 1996). The redshift 
distribution of the faintest galaxies in the survey extends out to a redshift of
about 0.2. Saunders et al.\ (1990) and Oliver et al.\ (1992) found that pure 
luminosity evolution of the form $(1+z)^{3.1}$ or pure density evolution of the 
form $(1+z)^{6.7}$ can account for the results. The inferred form of luminosity
evolution is comparable with that determined for radio-galaxies, quasars and
optically-selected star-forming galaxies (see e.g. Dunlop \& Peacock 1990; 
Hewett, Foltz \& Chaffee 1993; Lilly et al.\ 1996; Boyle \& Terlevich 1998 and 
Dunlop 1998). However, it is possible to fit the counts adequately by a 
range of evolution parameters if the typical dust temperature is varied: see 
Section\,3.3.

The mid- and far-infrared counts of more distant galaxies are being determined 
from the results of surveys using the {\it ISO} satellite; for example 
the observations of Kawara et al.\ (1997) and Lagache et al. (1998) at a 
wavelength of 175\,$\mu$m. At shorter wavelengths of 7 and 15\,$\mu$m, 
Oliver et al.\ (1997) and Hammer \& Flores (1998) derived counts 
from {\it ISO} observations of the {\it Hubble Deep Field} and CFRS fields 
respectively. However, it is difficult to translate between such a measurement of 
hot dust and the cooler dust that dominates the energy emission of a galaxy, 
and so the inferred global SFR (Rowan-Robinson et al.\ 1997) is very uncertain 
(Hughes 1996). Spectral features in dust emission in the mid-infrared waveband 
are also likely to affect the results (Xu et al. 1998). Much more information about 
the population of mid-infrared-luminous galaxies will be proved in 1999 by the 
{\it Wide-Field Infrared Explorer (WIRE)} satellite (Shupe et al. 1998). 

\subsection{The intensity of background radiation} 

The intensity of diffuse background radiation in the millimetre, submillimetre and 
far-infrared wavebands has been constrained by observations of the whole sky 
using the FIRAS (Fixsen et al.\ 1994) and DIRBE (Hauser et al.\ 1996) instruments 
aboard the {\it COBE} satellite (Puget et al.\ 1996; Guiderdoni et al. 1997; 
Fixsen et al. 1998; Hauser et al.\ 1998; Schlegel et al.\ 1998). 
At wavelengths longer than about 
400\,$\mu$m the cosmic microwave background radiation (CMBR) dominates the 
spectrum detected by FIRAS. After subtracting the CMBR signal and the
emission from Galactic dust, the isotropic signal from extragalactic sources 
remains. This subtraction is a complex process, and will improve in the
future. The {\it Planck Surveyor} satellite (Bersanelli et al.\ 1996) will also obtain 
much more accurate maps of the submillimetre-wave sky. However, our 
knowledge of the intensity of extragalactic background radiation in the 
millimetre/submillimetre and far-infrared wavebands is now substantially 
complete (see Fig.\,2). Consistent limits to the background radiation intensity at 
shorter wavelengths have been determined by Dwek \& Slavin (1994), Biller 
et al.\ (1998) and Dwek et al.\ (1998). 
The energy-normalized background spectrum $\nu I_\nu$ hence has a peak at a 
wavelength between about 100 and 200\,$\mu$m.

\begin{figure}
\begin{center}
\epsfig{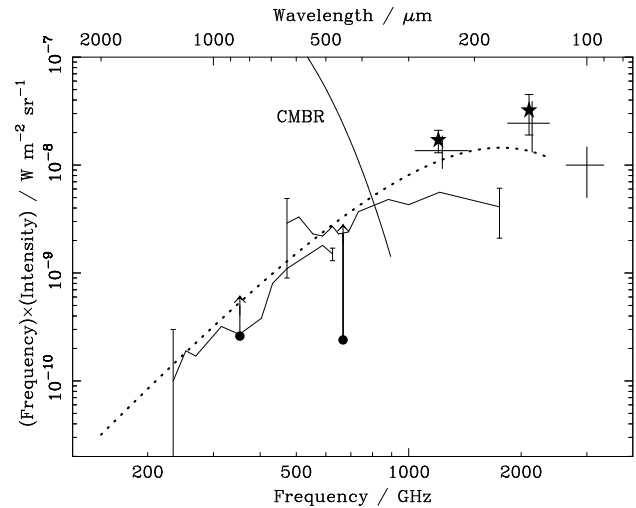}
\end{center}
\caption{The intensity of background radiation in the millimetre, submillimetre 
and far-infrared wavebands, as inferred by: Puget et al.\ (1996) -- thin solid lines 
with error bars at the ends; Fixsen et al.\ (1998) -- dotted line; Schlegel et al.\ 
(1998) -- stars; Hauser et al.\ (1998) and Dwek et al.\ (1998) -- thick solid 
crosses. The filled circles represent lower limits to the background intensity 
inferred from the counts of SIB. The connected arrows point to the 
corresponding background intensity that is extrapolated using the count model 
described in SIB and Blain, Ivison \& Smail (1998a, hereafter BIS). 
}
\end{figure}

\subsection{Millimetre/submillimetre-wave source counts}

Millimetre/submillimetre-wave observations using both single-antenna 
telescopes and interferometer arrays can be used to detect and limit the 
counts of discrete sources. At a wavelength of 2.8\,mm, Wilner \& Wright (1997) 
used the BIMA interferometer to image the Hubble Deep Field to a 1$\sigma$ 
sensitivity of 0.71\,mJy. No sources were detected at a significance greater than 
5$\sigma$, which translates into a 3$\sigma$ upper limit to the counts of 
160\,deg$^{-2}$ at a flux density of 2\,mJy. 

At 850\,$\mu$m SIB detected six 
sources, which corresponds to a surface density of 
$(2.5 \pm 1.4) \times 10^3$\,deg$^{-2}$ at a flux density of 4\,mJy. The brightest 
850-$\mu$m source (Ivison et al.\ 1998b) was also detected at a wavelength of 
450\,$\mu$m, indicating a very weak 3$\sigma$ limit of about $10^3$\,deg$^{-2}$ 
to the 450-$\mu$m counts at a flux density of 80\,mJy. Observations of five 
additional clusters (Smail et al.\ in preparation) have now extended the area 
of this survey to greater than 0.01\,deg$^2$. 
The surface density of discrete submillimetre-wave 
sources detected in the complete survey is fully consistent with measurements of 
the background radiation intensity discussed above (SIB). In deep images of 
dusty disks around stars, Holland et al.\ (1998) detected four unknown, probably 
extragalactic objects, indicating a count of $(10\pm6)\times10^2$\,deg$^{-2}$ at 
a flux density of 8\,mJy. At shorter wavelengths, Phillipps (1997) presented a 
350-$\mu$m 
8.6-arcmin$^2$ map made using the SHARC bolometer array at the Caltech 
Submillimetre Observatory, which contains no significant sources and reaches 
a 1$\sigma$ sensitivity of about 100\,mJy.

The results of blank-field surveys in the {\it Hubble Deep Field} 
(Hughes et al.\ 1998), CFRS fields (Eales et al.\ 1998) and 
Hawaii Deep Survey fields (Barger et al.\ 1998), and independent observations of 
clusters will reduce the errors on the 850-$\mu$m counts (see Fig.\,11c for
more details). The large complete cluster survey of Smail et al.\ (in preparation) 
will probably yield the most accurate constraints; the source counts derived from 
this large sample are discussed in a forthcoming paper (Blain et al.\ submitted). 

\begin{figure}
\begin{center}
\epsfig{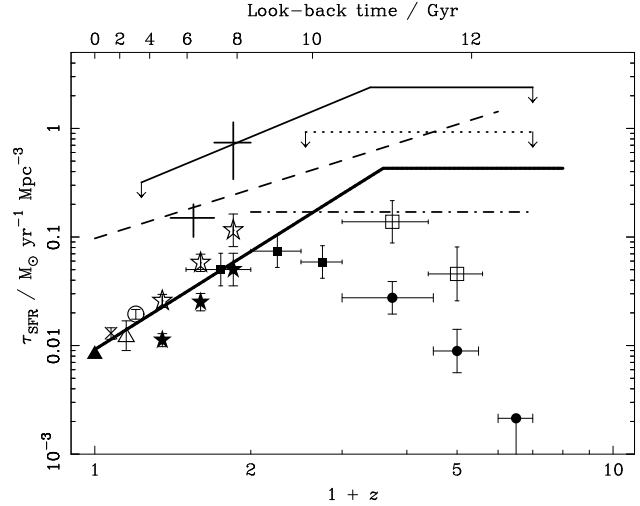}
\end{center}
\caption{
Histories of star formation derived from observations of dust emission and metal
abundance in clusters. The data points from Fig.\,1 are replotted for comparison. 
Blain \& Longair (1993b) -- dashed line -- and Burigana et al.\ (1997) -- solid 
and dotted lines with arrows -- imposed limits from background radiation
 intensities. The SFR inferred by Rowan-Robinson et al.\ (1997) and Hammer 
\& Flores (1998) based on {\it ISO} observations at 7 and 15\,$\mu$m are shown 
by the solid crosses and empty stars respectively. The star-formation history 
associated with the count model discussed in SIB and BIS is shown by the 
thick solid line. The dot-dashed line shows the estimate derived by Mushotzky
\& Loewenstein (1997) based on the metallicity of intracluster gas.
}
\end{figure}

\subsection{Early results} 

Measurements of the far-infrared background radiation intensity, and deep 
{\it ISO} observations in the mid-infrared waveband have led to estimates of the 
star-formation history that is obscured by dust, as shown in Fig.\,3. The 
comoving luminosity density emitted by dust in the far-infrared waveband 
$\rho_{\rm FIR}$ can be related to the density of metals generated in high-mass 
stars by the present epoch $\Omega_{\rm m}(0)$ in a population of galaxies that 
undergo their first generation of star formation at a redshift $z_0$, by,
\begin{equation}
\rho_{\rm FIR}(z) = 0.007\bar\rho c^2 
{
{H_0\Omega_{\rm m}(0)} \over { {\rm ln} (1+z_0)} 
}
(1+z)^{3/2}, 
\end{equation}
in an hierarchical model (Blain \& Longair 1993b). The smoothed density of
the Universe is $\bar\rho$. The relationship is similar in 
other models. The background radiation intensity shown in Fig.\,2 can be 
explained by the generation of about $\Omega_{\rm m}=10^{-3}$ since a redshift 
$z_0=5$ in this model. The estimated star-formation rate, assuming a standard 
conversion factor between far-infrared luminosity and the SFR (see Section\,4) 
provides the star formation history shown in Fig.\,3. The limits derived in a 
similar analysis by Burigana et al.\ (1997) are also shown. Counts of galaxies at 7 
and 15\,$\mu$m can be used to infer a star-formation rate at $z\ls 1$ 
(Rowan-Robinson et al.\ 1997; Hammer \& Flores 1998), although potentially 
with a very large uncertainty (see Fig.\,3).

\section{Determining the history of obscured star formation}

\subsection{Strategy}

Even before the completion of redshift surveys of submillimetre-selected
galaxies we can still obtain useful constraints on the star-formation history at 
high redshifts by combining
observations of discrete sources with measurements of the intensity of 
background radiation in the millimetre, submillimetre and far-infrared 
wavebands to derive self-consistent models for the  
evolution of the population of dusty galaxies. At very low redshifts, $z \le 0.2$, 
these are constrained to match the normalization and evolution of the 
60-$\mu$m counts of {\it IRAS} galaxies. Note that redshifts less than two are 
described as `low' in this paper. 

First, an estimate of the typical dust temperature and the slope of the evolution 
function at low redshifts is obtained by fitting a pair of functional forms, which 
asymptote to power laws in $(1+z)$ at low redshift, to both the 60- and 
175-$\mu$m counts. Most of the degeneracy between the values of dust 
temperature and evolution parameter is broken by including information from the 
175-$\mu$m counts, and so a robust estimate of the form of evolution of dusty 
galaxies at low redshifts can be obtained.

Secondly, the resulting models of the evolving 
population of dusty galaxies are extrapolated out to high redshifts, and are
constrained to fit the background radiation spectrum and 
submillimetre-wave counts. 

Thirdly, we employ a simple chemical evolution 
model to check whether or not the best-fitting models are self-consistent. A 
sufficiently large mass of dust must be generated in each model to absorb early 
starlight and to account for the large amounts of far-infrared emission
(Eales \& Edmunds 1996, 1997; Frayer \& Brown 1997), subject to the constraint 
on the temperature of emitting dust derived at low redshifts. The total mass of 
heavy elements 
produced in high-mass stars throughout the history of the Universe 
(Savage \& Sembach 1996) is also checked for consistency with observations.

In order to avoid introducing a large number of free parameters, we aim to explain
all the data using a single population of dusty galaxies, described by the 
60-$\mu$m luminosity function, with a fixed dust temperature $T_{\rm d}$ and 
emissivity $k_{\rm d}$. Saunders et al.\ (1990) fit the 60-$\mu$m counts using 
both a warm and cool component, and Pearson \& Rowan-Robinson (1996) 
adopt a similar scheme of cool `cirrus' and warm `starburst' galaxies.
Although distinct populations of dusty galaxies exist in the local Universe
(Sanders \& Mirabel 1996), in this paper we are investigating the global 
evolution of galaxies, driven by observations that are sensitive to galaxies out to 
redshifts of about 10. Hence, while such multi-component models may be 
required to explain more detailed observations in the future, we do not believe 
that the introduction of additional parameters is justified by the data at present. 
Here we fit four observations -- the evolution of the 60-$\mu$m count, the 175- 
and 
850-$\mu$m counts and the background radiation spectrum -- using a model 
that contains four parameters -- a typical dust temperature $T_{\rm d}$, a 
low-redshift slope of the evolution function $p$, and two redshifts, 
$z_{\rm max}$ and $z_0$, the redshifts beyond which evolution slows
down and at which the first galaxies appear respectively.
 
\subsection{Spectral energy distribution}

We assume that the frequency-dependent emissivity function of dust $k_\nu$ can 
be described in the millimetre and submillimetre wavebands by a simple 
power-law function $\nu^{1.5}$ (Hughes 1996; Ivison et al.\ 1998a). 
The dust emission 
spectrum is hence a modified Planck function with a Rayleigh--Jeans power-law 
index of $3.5$. At shorter far-infrared wavelengths this function would decline 
exponentially as a function of frequency, turning over at a wavelength of 
$130\pm50$\,$\mu$m for a typical dust temperature of order 
$50\pm20$\,K (Hughes 1996). A power-law spectrum of the form 
$\nu^{-\alpha}$ is grafted onto the modified Planck function at wavelengths 
shorter than that for which the equivalent power-law index of the modified 
Planck function is less than -$\alpha$. The slope of this short-wavelength 
spectrum reflects the fraction of dust grains in the model galaxy at 
temperatures larger than the defining temperature in the Planck function. A 
reasonable fit to radiative transfer models of dusty star-forming galaxies is
provided by a value $\alpha = 2.2$. These models are important for interpreting 
observations at wavelengths shorter than 50\,$\mu$m but not very relevant 
here (see Goldschmidt et al.\ 1997 and Xu et al. 1998).  

\begin{figure*}
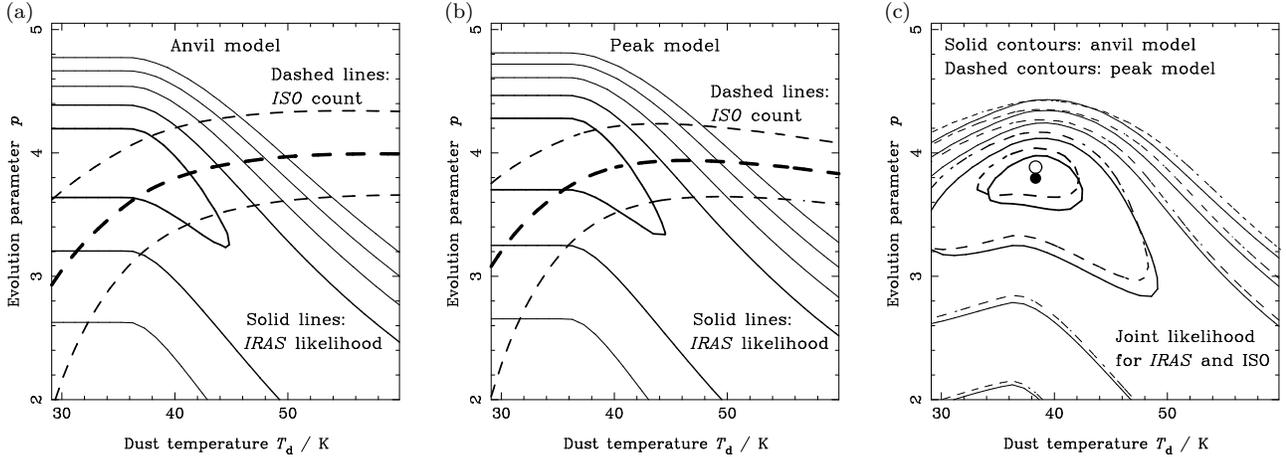

\begin{minipage}{170mm}
(a) \hskip 53.5mm (b) \hskip 53.5mm (c)
\begin{center}
\vskip -5mm
\epsfig{file=fig4a.ps, width=5.75cm, angle=-90} \hskip 5mm
\epsfig{file=fig4b.ps, width=5.75cm, angle=-90} \hskip 5mm
\epsfig{file=fig4c.ps, width=5.75cm, angle=-90}
\end{center}
\caption{The results of fitting the {\it IRAS} 60-$\mu$m counts and the {\it ISO} 
175-$\mu$m count (Kawara et al. 1997) using the anvil and peak models: 
equations (3) and(4) respectively. In (a) and (b) the solid contours show values of 
$\chi^2/N$ that are 1, 2, 3, 4 and 5$\sigma$ away from the local minimum $\chi^2$ 
values derived from the {\it IRAS} count data; the thick and thin dashed lines 
trace the locus of parameters for which the exact, half and twice the {\it ISO} 
count is reproduced respectively. Likelihood contours, representing the same 
$\chi^2$ values, obtained by fitting the models jointly to the {\it IRAS} and 
{\it ISO} counts are shown in (c). The best-fitting parameters in the anvil and 
peak model are represented by a solid and empty circle and correspond to 
$\chi^2$ values of 2.63 and 2.73 respectively. As the {\it IRAS} and {\it ISO} 
sources are expected to be at low redshifts, there is little difference between the 
predictions of the models.
}
\end{minipage}
\end{figure*}

\subsection{Fitting the low-redshift counts}

The 60-$\mu$m counts of {\it IRAS} galaxies can be used to constrain the form of 
evolution at redshifts $z\ll1$ (Saunders et al.\ 1990; Oliver et al.\ 1992). If the
luminosity function $\Phi(L, z)$ is parametrized using the two functions $n(z)$ 
and $g(z)$, which describe density and luminosity evolution respectively, then,   
\begin{equation}
\Phi(L,z) = n(z) \Phi\left[ {L\over{g(z)}}, 0 \right]. 
\end{equation}
Both $n(z) = (1+z)^{6.7}$ and $g(z)=1$ and $n(z) = 1$ and $g(z) = (1+z)^{3.1}$ 
were found to provide good fits to the counts. The first of these cases is pure 
density evolution, and the second is pure luminosity evolution. Only models of 
pure luminosity evolution are considered here, because it is impossible to 
reproduce the counts of distant submillimetre-selected objects in a model of 
pure density evolution, without overproducing the intensity of far-infrared 
background radiation and the density of heavy elements generated in 
nucleosynthesis by a huge amount: between 50 and 100 times. This is because 
the effects of pure density and pure luminosity evolution are almost the same in 
the calculation of the background intensity and heavy element abundance, but 
luminosity evolution has a more significant effect in the calculation of the counts. 
This is not to say that no density evolution occurs, but luminosity evolution 
must be dominant if the existing data is accurate. 

Two classes of evolution models are 
considered: an `anvil' model defined by the evolution function, 
\begin{equation}
g_{\rm a}(z) = \left\{ \begin{array}{ll}
(1+z)^p, & z \le z_{\rm max};\\
(1+z_{\rm max})^p, & z > z_{\rm max}, 
\end{array}\right.
\end{equation}
and a `peak' model, defined by the function, 
\begin{equation}
g_{\rm p}(z) = 
\displaystyle{ 2 
\left( 1 + \exp{ { z \over {z_{\rm max}} } } \right )^{-1} 
(1+z)^{p+(2z_{\rm max})^{-1}}. 
}
\end{equation}
In both models no galaxies exist beyond a redshift of $z_0$. At very low redshifts 
$g_{\rm p}$ has the asymptotic form $(1+z)^p$. 

These functional forms allow rather general and plausible evolution 
functions to be simply described: compare Cole et al.\ (1994). The peak model 
can be used to describe the star-formation history derived from optical and 
near-infrared observations in Fig.\,1 adequately: 
see Table\,1 and Fig.\,8. By requiring an evolving {\it IRAS} 60-$\mu$m 
luminosity function to fit both the form of evolution of the 60-$\mu$m counts and 
175-$\mu$m {\it ISO} count, best-fitting values of the dust temperature 
$T_{\rm d}$ and low-redshift evolution parameter $p$ can be determined, as 
shown in Fig.\,4. The derived best-fitting value of the dust temperature 
$T_{\rm d}=38\pm4$\,K in both the anvil and peak models. In the anvil model the 
best-fitting evolution parameter $p=3.8\pm0.2$, and in the 
peak model the best-fitting $p=3.9\pm0.2$. A slightly stronger form of evolution 
is required in the peak model because the exponential term reduces the strength 
of the evolution at redshifts less than $z_{\rm max}$. Note that these slopes are 
very similar to the value required to fit the 
rate of evolution of optically selected star-forming galaxies determined at a 
wavelength of 280\,nm by Lilly et al.\ (1996): they found $p=3.9\pm0.75$.
The estimated dust temperature appears to be consistent with the typical value 
of less than 40\,K derived for a sample of very distant quasars by Benford et al. 
(1998). The 60-$\mu$m counts predicted by the best-fitting models are compared 
with the {\it IRAS} data and the predictions of a non-evolving model in Fig.\,5.

Equally good fits to low-redshift evolution of dusty galaxies are provided by the 
peak and anvil models. However, the 60-$\mu$m counts provide no significant 
restrictions on the form of evolution at $z\gs 1$, and the {\it ISO} 
counts are not expected to include a significant proportion of galaxies at 
redshifts greater than 2. Hence, the submillimetre-wave counts and 
background intensities are required in order to investigate the history of star 
formation at earlier epochs. 

\subsection{Fitting the evolution of distant galaxies}

In order to impose limits to the form of evolution at high redshifts, the
peak and anvil models of pure luminosity evolution are used to predict both
the spectrum of submillimetre-wave background radiation and the counts of 
dusty galaxies at wavelengths of 2.8\,mm, 850 and 450\,$\mu$m. The density 
of metals generated in high-mass stars by the present epoch is also calculated. 
The models used to derive these quantities are discussed by Blain \& Longair 
(1993a, 1996). 

The results of these calculations are presented in Fig.\,6 for illustration in the
anvil model, assuming the values of $p$ and $T_{\rm d}$ derived above, as a 
function of $z_{\rm max}$ and $z_0$, the remaining free parameters in the 
evolution function (equation 3). Likelihood contours derived for independent fits 
to the background intensities reported by Fixsen et al.\ (1998), and a
combination of the results of Schlegel et al.\ (1998) and Hauser et al.\ (1998) are 
compared with the likelihood of a fit to SIB's 850-$\mu$m counts. 

Additional constraints can be imposed by considering source counts at 
wavelengths of 2.8\,mm (Wilner \& Wright 1997) and 450\,$\mu$m (SIB; Barger et
al. 1998), and the density parameter of heavy elements generated by high-mass 
stars by the present epoch $\Omega_{\rm m}(0)$, expected to be approximately 
$10^{-3}$. If solar metallicity (Savage \& Sembach 1996) is typical of the Universe 
as a whole, and the density parameter in baryons $\Omega_{\rm b}h^2=0.019$ 
(Burles \& Tytler 1998), then $\Omega_{\rm m}\simeq 1.9\times10^{-3}$. These
constraints are weaker than those imposed by the counts and background data 
and shown in Fig.\,6. The background estimate of Puget et al. (1996) provides a 
substantially weaker constraint on our models than any of the other estimates 
of the background and so does not add anything to our analysis. As noted by 
Fixsen et al.\ (1998), if the H$^+$ correction employed by Puget et al.\ (1996) is 
omitted, then the two determinations of the background intensity are in good 
agreement (see Fig.\,4b of Fixsen et al.\ 1998). The results of further analysis of 
the {\it COBE} data will probably increase the certainty with which the 
background is known, but, as shown in Fig.\,6 and discussed below, the newly 
determined submillimetre-wave counts are the crucial data for discriminating 
between models. 

In Fig.\,7 the joint likelihood for a fit to all the data, apart from Puget et al.'s 
background is presented for both the peak and anvil models. All the data
can be fitted very satisfactorily. Note that changing the values of the parameters 
$p$ and $T_{\rm d}$, which were determined using low-redshift data in 
Section\,3.3, within their specified errors causes the probability contours in 
$z_{\rm max}$--$z_0$ parameter space to move by less than 1$\sigma$ from the 
positions shown in Fig.\,7.

\begin{figure}
\begin{center}
\epsfig{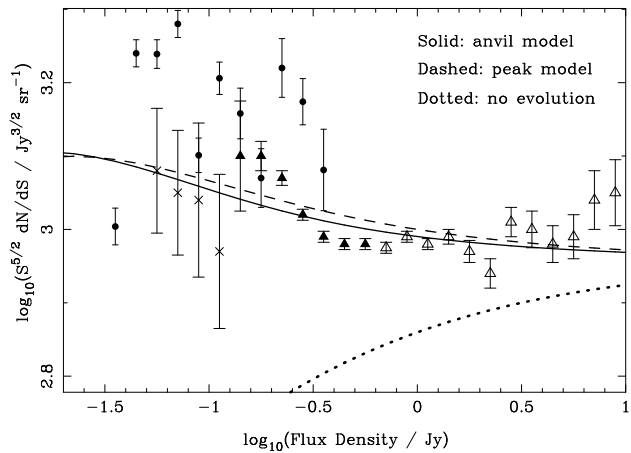}
\end{center}
\caption{Observed 60-$\mu$m counts of {\it IRAS} galaxies and the best-fitting 
peak and anvil models determined from the analysis in Section\,3.3, 
plotted in the format used by Oliver et al.\ (1992). For comparison, 
the counts expected in a non-evolving model are shown by the dotted line. The 
data are taken from Hacking \& Houck (1987; crosses), 
Rowan-Robinson et al.\ (1990; empty triangles), Saunders et al.\ (1990; filled 
triangles) and Gregorich et al.\ (1995; circles): see also Bertin et al. (1997).}
\end{figure} 

\begin{figure}
\begin{center}
\epsfig{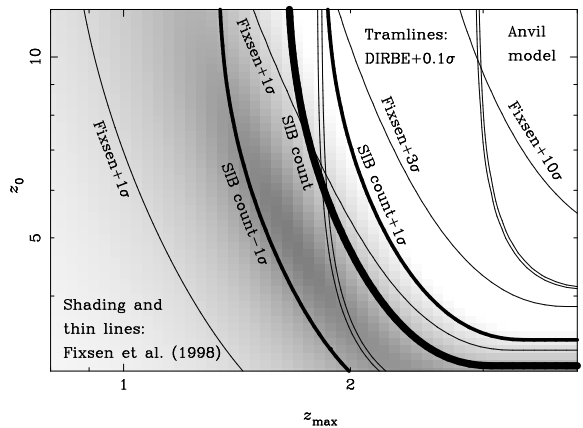} \hskip 5mm
\end{center}
\caption{The relative significance of the high-redshift constraints imposed on 
models of galaxy evolution that account for the low-redshift data. The results 
are derived in the anvil model for illustration. The corresponding diagram in 
the peak model has a very similar form. 
The results of fitting the 
FIRAS background radiation intensity determined by Fixsen et al.\ (1998), the 
DIRBE background (Schlegel et al.\ 1998; Hauser et al.\ 1998) and SIB's 
850-$\mu$m counts are compared. The counts are represented by the 
extremely thick solid line. The probability of fitting the FIRAS background is 
represented by the shaded regions; a minimum reduced $\chi^2$ value of 0.07 
is obtained for the best fit to this background. 
The probability of fitting the DIRBE background is represented by 
the tramlines.
}
\end{figure}

There are several interesting points to note in Fig.\,6. First, the regions of 
parameter space in which both the intensity of background radiation and the 
850-$\mu$m counts are fitted overlap satisfactorily. The counts and background 
estimates are consistent, although models that fit the counts are in better
agreement with the recent background estimates by Fixsen et al.\ (1998) 
than with Puget et al.\ (1996). Secondly, the tightness of the constraints imposed 
on the model parameters by the background measurements and the 850-$\mu$m 
counts are comparable. As the accuracy of the 850-$\mu$m counts is set to 
increase significantly when the complete cluster sample is presented (Blain et 
al.\ submitted; Smail et al. in preparation) and the data from other surveys 
becomes available, knowledge derived 
from the counts will be the determining factor in discriminating between models 
of galaxy evolution. Thirdly, the locus of the best-fitting parameters is a line 
and not a distinct peak. Hence, given the existing data, no individual model 
provides a best fit, but rather families of acceptable models trace 
out this locus (see Fig.\,7). In the final section we discuss how
we can discriminate  between members of these families. 

Three values of $z_0$ and $z_{\rm max}$ are selected in each model, highlighted 
by dots in Fig.\,7. The models defined by these sets of parameters are 
consistent with all the current submillimetre-wave data, and we list their
parameters in Table\,1. 

Submillimetre-wave observations are very sensitive to the form of galaxy 
evolution at high redshifts, and so provide a series of interesting 
cosmological tests (Blain 1998a). At present we are not prepared to use the 
limited available data to infer values of $\Omega_0$ and $\Omega_\Lambda$; 
however, note that low values of $\Omega_0$ or a non-zero value of 
$\Omega_\Lambda$ tend to increase the background radiation intensity for a 
given count. In an Einstein--de Sitter model, an exact fit to the counts requires 
a more significant form of evolution than an exact fit to the background (see 
Fig.\,6). Hence, at a first level of approximation, a high-density 
Universe would seem to be more favoured by this data. We will return to this 
issue once more accurate counts have been determined.

\subsection{An optically-motivated model}

The likelihood contours in Fig.\,7 have been used to select six best-fitting peak 
and anvil models, to test the robustness of the procedure, and to investigate 
whether or not observational tests can be devised to distinguish between the 
different cases. However, these best-fitting models have been selected on the 
basis of observations in the far-infrared and submillimetre wavebands alone. If 
the extinction corrections applied to account for reddening in faint 
Lyman-dropout selected sources (Pettini et al.\ 1998a,b) are correct, then the 
empty squares in Fig.\,1 represent the global SFR in a high-redshift population of 
non-active sources, with individual SFRs in the range 10 to 
50\,M$_\odot$\,yr$^{-1}$. The effects of this population, which consists of less 
luminous sources than the galaxies detected by SIB, should be included in the 
models. It is possible to use a peak model -- Peak-G in Table\,1 -- to connect 
this population and the low-redshift {\it IRAS} galaxies, assuming that they are 
subject to the same form of low-redshift evolution as the {\it IRAS} galaxies. The 
choice of $z_0$ in this model is unimportant, as the SFR at $z>5$ is effectively 
zero in this model. 

In order to explain the high-redshift data -- the submillimetre-wave background 
radiation intensity and counts, and the majority of the 175-$\mu$m {\it ISO} 
counts -- an additional second population of more luminous, but obscured, 
galaxies must be included. This 
`Gaussian' population is based on the {\it IRAS} luminosity function, and is
assumed to undergo pure luminosity evolution of the form, 
\begin{equation}
g_{\rm g}(z) = \Theta \exp{ \left\{ - { { \left[ t(z)-t(z_{\rm p})\right]^2 } \over 
{2 \sigma^2} } \right\} }.   
\end{equation}
This function represents a burst in the SFR of the population at a characteristic 
epoch, centred on the cosmic epoch that corresponds to redshift $z_{\rm p}$, 
with a characteristic timescale $\sigma$ and a normalization $\Theta$ with 
respect to the low-redshift {\it IRAS} luminosity function. To 
avoid introducing extra parameters, the dust temperature in this additional 
population remains at 38\,K. The three parameters $\Theta, z_{\rm p}$ and
$\sigma$ are determined by fitting the model to the submillimetre-wave counts 
and background radiation intensity (see Table\,1). 
Reasonably, $\Theta \sigma$, which is proportional to the total amount 
of star formation activity in the burst, is constrained more tightly than the 
individual values of $\Theta$ and $\sigma$: $\Theta \sigma=76\pm8$\,Gyr.

\begin{figure*}
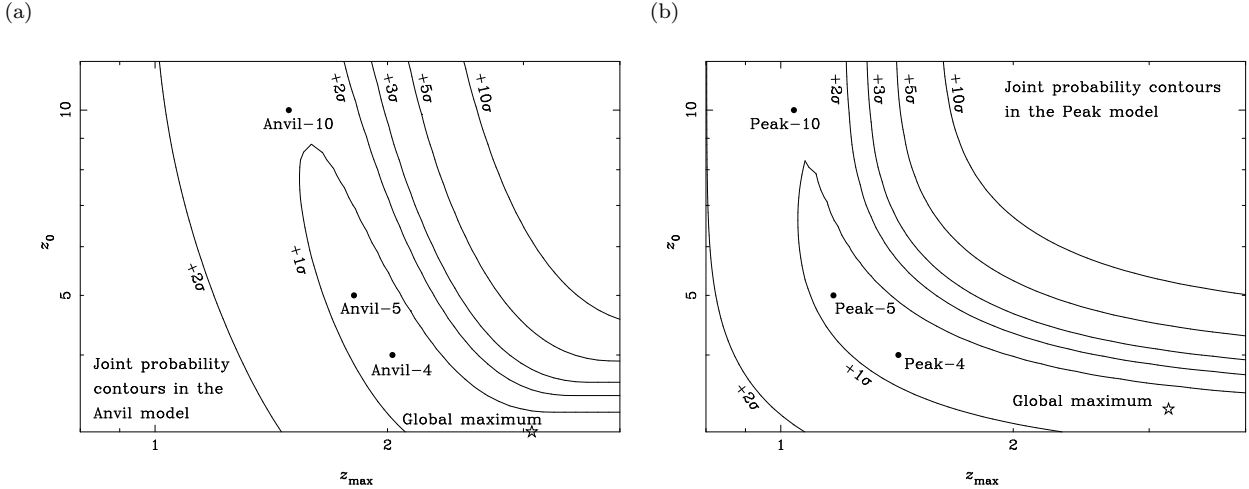

\begin{minipage}{170mm}
(a) \hskip 81mm (b)
\begin{center}
\epsfig{file=fig7a.ps, width=5.65cm, angle=-90} \hskip 5mm
\epsfig{file=fig7b.ps, width=5.65cm, angle=-90}
\end{center}
\caption{The joint probability that parameters defining the anvil and peak models 
fit to all the high-redshift data listed in Section\,3.4, apart from the background 
intensity of Puget et al. (1996). The maximum-likelihood values, marked by stars, 
correspond to $\chi^2$ values of 1.15 and 1.12 in (a) and (b) respectively. The 
chosen parameters in the models used in subsequent sections (Table\,1) are 
indicated by the labelled dots.
.}
\end{minipage}
\end{figure*}

\begin{table*}
\caption{The parameters defining a series of models that 
adequately describe the star formation histories derived from optical surveys 
and from the submillimetre-derived likelihood contours shown in Fig.\,7. The 
Peak-M and Peak-P 
models fit the optically-derived histories: see Fig.\,1. No extinction correction is 
included in the Peak-M model (Madau et al.\ 1996). The Peak-P model 
includes an extinction correction (Pettini et al.\ 1998). The Peak-G model is
chosen to fit both the low-redshift evolution and normalisation of the luminosity 
density of {\it IRAS} galaxies and the extinction-corrected optical points 
(Pettini et al.\ 1998a,b): see Section\,3.5. The other models fit all the 
submillimetre/far-infrared data. The line styles and thicknesses listed refer to the
form in which the results of these models are presented in Figs 8 to 13. 
$T_{\rm d}=38\pm4$\,K in all models. 
}
{\vskip 0.75mm}
\hrule{\vskip 1.2mm}
\begin{tabular}{ p{1.2cm} p{0.6cm} p{1.4cm} p{1.4cm} 
p{0.9cm} p{2.1cm} p{1.5cm} p{1.5cm} p{3.7cm}}
Model & $z_0$ & $z_{\rm max}$ & $p$ & $\Omega_{\rm m}$ / & 
$\tau_{\rm SFR}(0)$ / $10^{-3}$ & Line style & Line & Comments\\
 & & & & 10$^{-3}$ & M$_\odot$\,yr$^{-1}$\,Mpc$^{-3}$ & & Thickness & \\
\end{tabular}
{\vskip 1.2mm}
\hrule
{\vskip 2.7mm}
\begin{tabular}{ p{1.2cm} p{0.6cm} p{1.4cm} p{1.4cm} 
p{0.9cm} p{2.1cm} p{1.5cm} p{1.5cm} p{3.7cm}}
Peak-M & 10 & $0.39\pm0.02$ & $4.6\pm0.3$ & 0.16 & $8.0\pm0.8$ & 
Dot-dashed & Thin & Fits Madau et al.\ (1996) \\
Peak-P & 10 & $0.48\pm0.05$ & $4.9\pm0.4$ & 0.31 & $7.2\pm0.8$ & 
Dot-dashed & Thick & Fits Pettini et al.\ (1998a,b) \\
\noalign{\vskip 1.5mm}
Anvil-10 & 10 & $1.49\pm0.15$ & $3.8\pm0.2$ & 0.85 & 7.7 & Solid & Thick & \\
Anvil-5 & 5 & $1.81\pm0.15$ & $3.8\pm0.2$ & 0.99 & 7.7 & Solid & Medium & \\
Anvil-4 & 4 & $2.03\pm0.15$ & $3.8\pm0.2$ & 1.05 & 7.7 & Solid & Thin & \\
\noalign{\vskip 1.5mm}
Peak-10 & 10 & $1.04\pm0.15$ & $3.9\pm0.2$ & 0.79 & 7.7 & Dashed & Thick & \\
Peak-5 & 5 & $1.17\pm0.15$ & $3.9\pm0.2$ & 0.84 & 7.7 & Dashed & Medium & \\
Peak-4 & 4 & $1.42\pm0.15$ & $3.9\pm0.2$ & 0.90 & 7.7 & Dashed & Thin & \\
\noalign{\vskip 1.5mm}
Peak-G & 10 & $0.63\pm0.04$ & $3.9\pm0.1$ & 0.35 & 7.7 & Dotted & Thin & 
Base for Gaussian model\\
 \noalign{\vskip 1.5mm}
Gaussian & 10 & N/A & N/A & 0.91 & N/A & Dotted & Thick & 
Always added to above\\
& & & & & & & &$z_{\rm p}=2.1\pm0.2$ \\
& & & & & & & &$\sigma=0.8\pm0.5$\,Gyr \\
& & & & & & & &$\Theta=95\pm40$ \\ 
\end{tabular}
{\vskip 1.2mm}
\hrule
\end{table*}

\section{Consistency arguments} 

\subsection{Generation of dust and stars}

We have adopted a single-temperature dust model to evaluate a star formation 
history in obscured galaxies that is consistent with all existing observations at
wavelengths between the millimetre and mid-infrared wavebands. The
temperature was obtained in Section\,3.3 using the low-redshift count data. 
For these models to be self-consistent, a sufficiently large mass of dust 
must be present at each epoch to absorb the optical/ultraviolet continuum 
luminosity of young star-forming regions and re-emit it at the required 
temperature. This type of condition was discussed in the context of a starburst 
in an individual galaxy by Eales \& Edmunds (1996, 1997). While preparing 
previous papers (Blain \& Longair 1993a, 1996), the results of 
`back-of-the-envelope' calculations suggested that this effect would not be 
particularly significant, especially as there was so little data with which to 
constrain model predictions. However, now that much more data is available, it is 
important to confirm the self-consistency of our models. 

The relationship between the mass, temperature and total bolometric luminosity
of dust grains is reasonably well determined. If the quantities are all expressed in 
densities per unit comoving volume, then the luminosity density emitted in the
rest-frame far-infrared waveband
\begin{equation}
\left( { { \rho_{\rm FIR} } \over { {\rm L}_\odot\,{\rm Mpc}^{-3} } } \right) 
\simeq 
2 \times 10^{-5} 
\left( { { \rho_{\rm d} } \over { {\rm M}_\odot\,{\rm Mpc}^{-3} } } \right) 
\left( { {T_{\rm d}} \over {\rm K} } \right)^{5.5}, 
\end{equation}
where $\rho_{\rm d}$ is the comoving density of dust grains, and a $\nu^{1.5}$ 
emissivity law is assumed. Regardless of whether the dust is distributed
uniformly or in clumps, equation (6) implies that there is a well-defined minimum 
temperature that the dust must have in order to emit a given luminosity density. 
Because of the strong temperature dependence on the right-hand side, it is 
possible for a small fraction of the dust at a higher than average temperature to 
dominate the emission of energy; for example, the
same luminosity is emitted by a mass of dust at a uniform temperature as by 
only 1 per cent of the grains at a temperature increased by a factor of 
2.3. However, it is not possible for a required luminosity to be generated at a
uniform lower temperature.

For an obscured star formation model to be self-consistent, enough 
dust must be generated by prior generations of stars to satisfy the constraint 
on dust temperature imposed by equation (6) at a temperature of 
38\,K. If this condition is not satisfied, then insufficient absorption can take place 
at high redshifts. By rearranging equation (6), the limit is given by 
\begin{equation} 
\left( { {T_{\rm d}} \over {\rm K} }\right) > 7.1
\left( { { \rho_{\rm FIR} } \over { {\rm L}_\odot\,{\rm Mpc}^{-3} } } \right)^{0.182}
\left( { { \rho_{\rm d} } \over { {\rm M}_\odot\,{\rm Mpc}^{-3} } } \right)^{-0.182}.
\end{equation}

$\rho_{\rm FIR}$ is expected to be directly proportional to the SFR per unit 
comoving volume $\rho_{\rm SFR}$; thus
\begin{equation}
\rho_{\rm FIR} = K \rho_{\rm SFR}. 
\end{equation}
Thronson \& Telesco (1986) derived a value of $K = 1.5 \times
10^{9}$\,L$_\odot$\,M$_\odot^{-1}$\,yr for the entire range of stellar masses
included in a Salpeter initial mass function (IMF).   We note that
the choice of this IMF is somewhat arbitrary at this time owing to our
lack of knowledge of the IMF in galaxies in the early Universe.
Rowan-Robinson et al.\ (1997) 
suggest a value of $K= 2.2 \times 10^{9}$\,L$_\odot$\,M$_\odot^{-1}$\,yr,
and we assume this value here. If only OBA stars are formed, 
then the value of $K$ is expected to increase by a factor of about 3. Equations (6) 
and (8) provide a simple relationship between the bolometric luminosity emitted 
by dust and the comoving density of stars formed. Hence, given any 
star formation history, the total density of material converted into stars by any 
epoch 
\begin{equation}
\rho^*(z) = K^{-1} \int_{t(z)}^{t(z_0)} \rho_{\rm FIR} \, {\rm d}t + \rho^*_{\rm III}, 
\end{equation}
can be calculated. $\rho^*_{\rm III}$ is the comoving density of stars formed in any
Population III component at redshifts greater than $z_0$.

The evolution of the density of dust, which is required to evaluate the limit in 
equation (7), can also be estimated using a simple model. Throughout 
the process of galaxy formation, dust is created as stars die and 
redistribute metals into the interstellar medium (ISM). 
Dust is also expected to be 
destroyed either by being bound up in newly formed stars or evaporated by the 
intense ultraviolet/blue light from OB stars. In general, both processes would be 
expected to proceed at a rate approximately following the SFR, and so it seems 
reasonable to assume that
\begin{equation}
\rho_{\rm d}(z) \simeq \epsilon_{\rm d} \rho^*(z), 
\end{equation}
in which $\epsilon_{\rm d}$ is the efficiency of dust formation: that is, if a unit 
mass of gas is converted into stars, then a fraction $\epsilon_{\rm d}$ of this 
mass is returned to the ISM in the form of dust grains after a short time. 
The details of the processing of elements in stars and the retention of metals in 
stellar remnants are extremely complex, but an efficiency of between $10^{-3}$ 
and $10^{-2}$ seems reasonable. If we assume that 30\,per 
cent of the material in stars with masses greater than 20\,M$_\odot$ is converted 
into metals and returned to the ISM, then for either a Salpeter IMF, with mass 
limits of 0.07 and 100\,M$_\odot$, or a Kennicutt IMF (Kennicutt 1983), with
mass limits of 0.1 and 100\,M$_\odot$, 2\,per cent of the material processed into 
stars is returned to the ISM in the form of metals. Note 
that changing the density of dust in
equation (7) by an order of magnitude modifies the lower limit to the dust 
temperature by only a factor of 1.5. 

\begin{figure*}
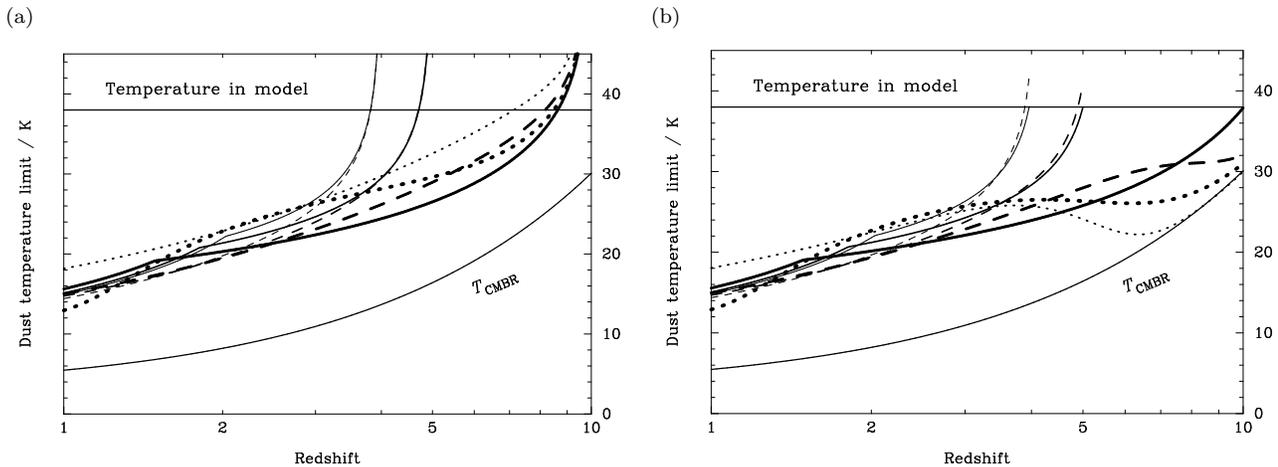

\begin{minipage}{170mm}
(a) \hskip 81mm (b)
\begin{center}
\vskip -2mm
\epsfig{file=fig8a.ps, width=5.5cm, angle=-90} \hskip 5mm
\epsfig{file=fig8b.ps, width=5.5cm, angle=-90}
\end{center}
\caption{The lower limits to the dust temperature imposed as a
function of redshift by the consistency argument detailed in Section\,4. The
temperature of 38\,K assumed in the models and the lower limit imposed by the 
CMBR temperature $T_{\rm CMBR}$ are also shown. In a self-consistent model, 
the lower limit to the temperature must be less than 38\,K at all epochs. 
The line styles and
thicknesses are interpreted in Table\,1. In (a) no Population-III stars are included. 
In (b) an additional 2 per cent of the total amount of star formation taking place 
in each model is added to generate dust in a Population III at redshifts greater 
than 10. When this additional small population is included, the models are all 
self-consistent.
}
\end{minipage}
\end{figure*}

Hence, by combining equations (7), (8), (9) and (10) a lower limit to the dust
temperature that is required for a self-consistent model can be derived. If this 
lower limit is less than the redshifted temperature of the CMBR 
$T_{\rm CMBR}(1+z)$, then clearly the limit is not a practical restriction 
(Eales \& Edmunds 1996). If the lower limit to the dust temperature is calculated 
assuming that the dust grains are in thermal equilibrium with the CMBR, then a 
dust temperature of 38\,K at low redshifts would correspond to a slightly higher 
temperature of 40\,K at $z=10$.

\subsection{The temperature limit for selected models}

The redshift-dependence of the dust temperature limit calculated for all seven  
models listed in Table\,1 that are fitted to the submillimetre-wave data is 
shown in Fig.\,8. In each case $\epsilon_{\rm d} = 10^{-2}$. In Fig.\,8(a) 
$\rho^*_{\rm III}=0$. With no Population III stars to generate early dust, the 
lower limit to the temperature exceeds the assumed value of 38\,K at redshifts 
close to $z_0$ in all the models. The consistency breaks down only at redshifts 
greater than about 8 for the models with $z_0=10$. Only a small fraction -- less
than 10 per cent -- of sources in these models are at redshifts greater than 8, as
shown by the redshift distributions in Fig.\,13. Hence, even this violation of 
consistency does not invalidate the conclusions of the paper. In Fig.\,8(b) a small 
Population III is included, in which $\rho^*_{\rm III}=10^{-2}\rho^*(0)$. By seeding 
the high-redshift gas with heavy elements and dust, this population ensures that
all the models are self-consistent.

The self-consistency condition is significant only if a low typical temperature
of dust, such as 38\,K, is assumed. If a significant fraction of submillimetre-wave 
sources are heated by AGN rather than high-mass stars, then systematically less 
dust forms by any epoch, and so the condition would be more difficult to satisfy. 
In an hierarchical model, a certain fraction of star formation is likely to occur in 
brief bursts at the epochs of mergers, and so only a fraction of the total mass of 
dust will be emitting strongly at any epoch. The lower limit to the dust 
temperature imposed by equation (7) would also increase in this case. In the
models of Baugh et al. (1998) less than 20\,per cent of stars form in such bursts, 
but the bursts could be responsible for a larger percentage of submillimetre-wave
sources. Either a large population of dusty AGN or hierarchical mergers 
at high redshifts would make it more difficult to obtain a self-consistent model.

Based on the calculations discussed above all the models listed in Table\,1 are 
self-consistent if a few per cent of star-formation activity occurs in Population III
stars. If this population does not exist, then the first generation of stars 
would not be obscured by dust for a brief period. Nevertheless, even in this 
case the results in this paper would not be modified by more than about 10\,per 
cent. 

\section{Results}

\subsection{Introduction}

Here we use the models presented in Section\,3 that account for all the 
available data in the millimetre/submillimetre and far-infrared wavebands to 
estimate the star formation history in obscured sources, and relate it to the 
observed consumption rate of neutral gas (Storrie-Lombardi, McMahon \& Irwin 
1996) and metal enrichment (Pettini et al.\ 1997) in Lyman-$\alpha$ absorbers. 

\subsection{The star formation history}

The star formation histories associated with the models presented in 
Table\,1 are plotted in Fig.\,9. Although there is a reasonable scatter in the 
results that fit the submillimetre-wave data, all the predictions lie within a factor 
of about 3 of each other over the critical range of redshifts from 2 to 5. The offset 
between the optically derived points and the submillimetre-derived curves is 
uncertain to within a factor of about 2. The estimated uncertainties in the 
values of the fitted parameters listed in Table\,1 can shift the curves through a 
region similar to that enveloped by the curves from model to model. The 
histories are fully consistent with limits derived from existing background 
radiation intensities (Blain \& Longair 1993b; Burigana et al.\ 1997). 

\begin{figure*}
\begin{minipage}{170mm}
\begin{center}
\epsfig{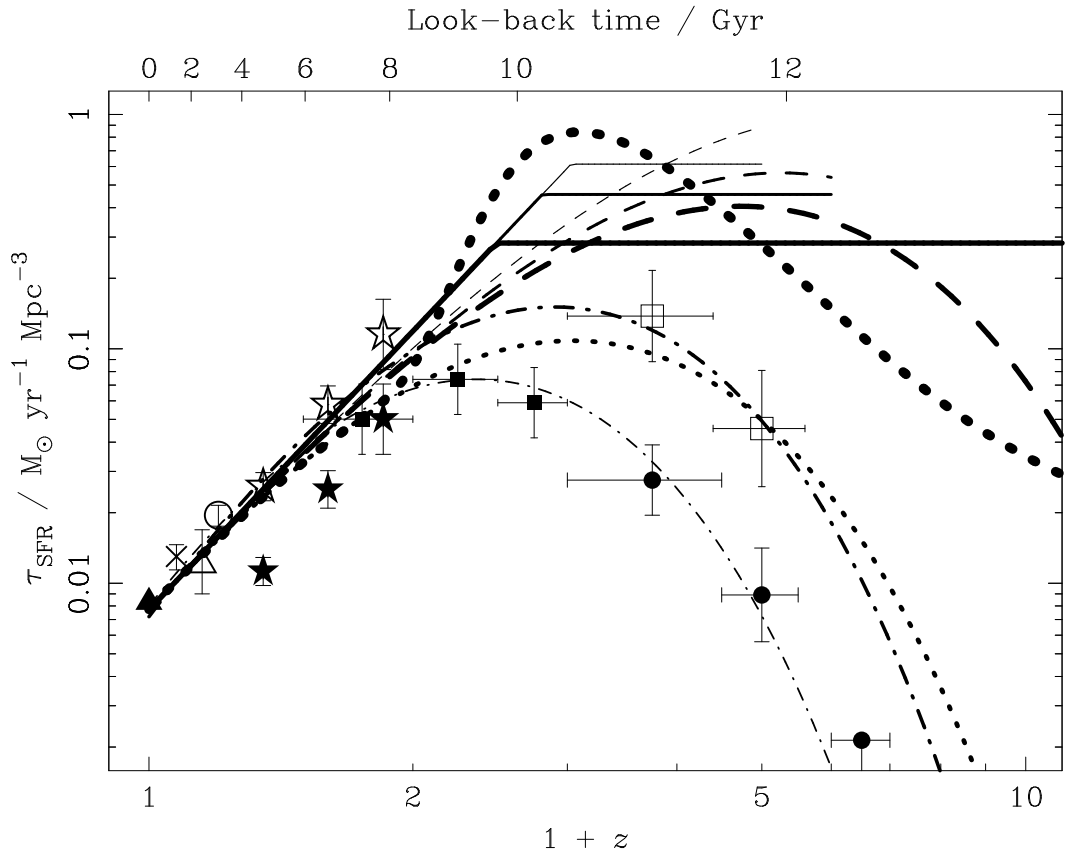}
\end{center}
\caption{Star formation histories in optically-selected and obscured galaxies 
presented for the models listed in Table\,1. The data points are identical to 
those plotted in Fig.\,3. The seven upper curves represent star formation 
histories that are consistent with all the available far-infrared and 
submillimetre-wave data; the three lower curves represent star formation 
histories that are consistent with the optical/ultraviolet/near-infrared data. 
}
\end{minipage}
\end{figure*}

At low redshifts the average slope of the star formation histories derived here is 
in good agreement with the form determined from observations of 
optically selected galaxies. However, there is a clear discrepancy between the 
two classes of history at high redshifts. The histories cannot be reconciled, 
even if the properties of the high-redshift, optically selected galaxies  
are corrected for the estimated dust extinction (Pettini et al.\ 1998a,b; 
Calzetti et al.\ 1995). 
A further correction to the dust extinction in the ultraviolet-selected samples
appears unlikely, and we therefore suggest that the remaining star formation
arises in a population of obscured galaxies which are either too red or too faint
in the rest frame ultraviolet waveband to be selected by the Lyman-dropout 
technique (Steidel et al.\ 1996b).   This suggestion can be tested by carrying out 
optical and near-infrared spectroscopy of submillimetre-selected sources to 
determine the fraction that lies at high redshifts $z\gs 3$.
 
The limited experience so far gained from follow-up observations of sources 
selected on the grounds of their powerful dust emission at more modest 
redshifts, {\it IRAS} F\,10214+4724 (Rowan-Robinson et al.\ 1991; 
Close et al.\ 1995) and SMM\,J02399$-$0136 (SIB; Ivison et al.\ 1998b), 
suggests that such sources can be detected and studied 
in the optical and near-infrared waveband in reasonable integration times. Hence 
the prospects for following up submillimetre-wave surveys and determining the 
typical properties of the optical counterparts appear to be quite good. Direct 
submillimetre-wave observations of the Lyman-break objects on which the 
optically derived star formation history is based will soon allow us to determine 
whether or not the extinction corrections inferred from their optical and 
near-infrared spectra are correct, and thus the contribution of such sources to 
the population of submillimetre-selected objects. 

Another population which may provide a link between the optical and 
submillimetre views of the distant Universe are the class of extremely red 
objects (EROs). These have been detected in deep multi-waveband surveys 
(Graham \& Dey 1996; Ivison et al.\ 1997; Cimatti et al.\ 1998; Dey et al.\ 1998), 
but not in sufficient numbers to be included in the derivation of the points in 
Fig.\,9. Nevertheless, detailed observations of these strongly obscured galaxies 
may help to unify the two regimes.

\subsection{Enrichment}

Observations of Lyman-$\alpha$ absorbers along the line of sight to distant 
quasars allow the evolution of the mass of neutral gas and 
the typical metallicity in the Universe to be traced as a function of epoch 
(Storrie-Lombardi et al.\ 1996; Pettini et al.\ 1997).

\subsubsection{Gas consumption and the density parameter in stars}

The evolution of the neutral gas content of the Universe has been determined 
by Storrie-Lombardi et al.\ (1996). The comoving density parameter of neutral gas 
$\Omega_{\rm g}$ appears to peak at a redshift of about 3. When combined with 
an estimate of the density parameter in the form of stars $\Omega_*(z)$, 
measured at the present epoch as $\Omega_*(0)=(5.9\pm2.3)\times10^{-3}$ 
(Gnedin \& Ostriker 1992), this result can be viewed in the context of the 
star formation history. In Fig.\,10(a) the mass of material that has been processed 
into stars is derived as a function of epoch using equation (9) for each of the 
star formation history models listed in Table\,1. 

The optically derived star formation history, as presented by Madau et al.\ (1996, 
1998) appears to be very consistent with the data presented in Fig.\,10(a). With or 
without dust corrections, that is in the Peak-M or Peak-P model respectively, 
$\Omega_*(z)$ builds up smoothly and $\Omega_* + \Omega_{\rm g}$ remains 
approximately constant as a function of epoch to within the errors. The same is 
true for the Peak-G model. However, the values of $\Omega_*(0)$ predicted by 
the submillimetre-derived star formation histories are considerably larger. The 
total density parameter of processed material exceeds the stellar density derived 
by Gnedin \& Ostriker (1992) by a factor of about 5. In fact, the values of 
$\Omega_*(0)$ determined in the submillimetre-derived models are comparable 
to the value of $\Omega_{\rm b}$. Note, however, that despite this large predicted 
stellar mass, about half of the total mass is expected to have been processed 
prior to a redshift of 2 to 3 in all the submillimetre-derived histories. The 
observed turn-over in the neutral gas fraction and the maximum rate of star 
formation in these histories are coincident in redshift.

If the background radiation intensity and counts determined in the submillimetre
waveband are not in error, then either a large fraction of the baryons in the 
Universe have been processed into stars, or the process of star formation in the 
distant sample must be different from that in nearby galaxies, with an initial IMF 
biased to high-mass stars.  A third possibility
is that  some fraction of the luminosity of distant dusty 
galaxies could be produced by AGN rather than by high-mass stars.

\begin{figure*}
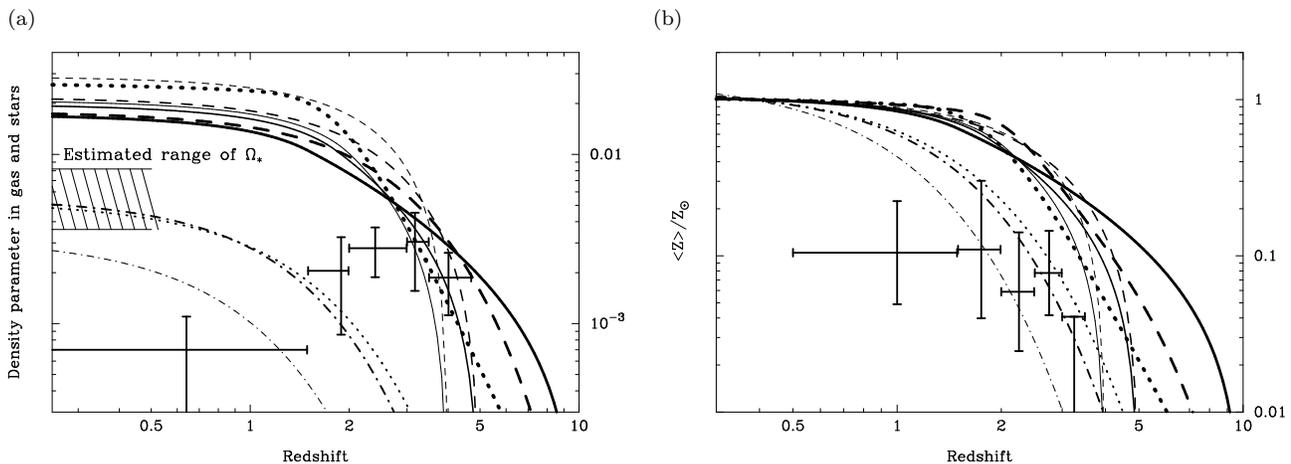

\begin{minipage}{170mm}
(a) \hskip 81mm (b)
\begin{center}
\vskip -2mm
\epsfig{file=fig10a.ps
, width=5.45cm, angle=-90} \hskip 5mm
\epsfig{file=fig10b.ps
, width=5.45cm, angle=-90}
\end{center}
\caption{(a) The density parameter of gas and stars as a function of redshift. The
curves represent the density parameter of material processed into stars in the 
models listed in Table\,1, assuming a Salpeter IMF with mass limits of 
0.07 and 100\,M$_\odot$. The data for the total density of stars at the present 
epoch (shaded region) were obtained by Gnedin \& Ostriker (1992). The data
points represent the density parameter in neutral hydrogen (Storrie-Lombardi
et~al.\ 1996). The lower three curves correspond to the models that fit the 
optically derived star formation histories listed in Table\,1. (b) The rate of 
increase of metallicity expected in the same models. The data points were 
determined by Pettini et al.\ (1997). 
}
\end{minipage}
\end{figure*}

First, let us consider the effects of modifying the IMF. If a Salpeter IMF is 
assumed, with a lower mass limit of 0.07\,M$_\odot$, then in all the models listed 
in Table\,1 about 65--70\,per cent of all stars formed are still burning at the 
present epoch. In the submillimetre-derived models this fraction is typically 
2--3\,per cent smaller than in the optically-derived models,  certainly not by 
enough to account for the factor of about 4 by which $\Omega_*$ exceeds the
observed value (Gnedin \& Ostriker 1992). If a shallower low-mass IMF is 
assumed, for example Kennicutt (1983), the fraction of stars still burning is 
reduced to about 40\,per cent. These results are unaffected by the 
value of the upper mass limit in the IMF. A population of extragalactic stars, not 
detected by Gnedin \& Ostriker, could account for some of the difference. 
In order to account for a factor of 2.5 discrepancy between the 
submillimetre-derived star formation histories and Gnedin \& Ostriker (1992) that 
remains after the fraction of 30\,per cent of stellar remnants is removed from 
the curves in Fig.\,10(a), a lower mass limit of about 0.7\,M$_\odot$ is required in 
the IMF. This is certainly a plausible scenario: a modified IMF in distant 
star-forming galaxies could account for these results, and would predict a stellar 
density at the present epoch for the submillimetre-derived models that is smaller 
by a factor of 4 compared with that shown in Fig.\,10(a). A lower mass limit to the
IMF of 3\,M$_\odot$ was suggested by Zepf \& Silk (1996) in order to explain 
mass-to-light ratios of elliptical galaxies, and Rieke et al. (1993) interpret 
observations of M82 using the same cutoff value. The potential 
variations in the high-redshift IMF are discussed by Larson (1998). 

Alternatively, we consider the effects of a population of obscured sources 
powered by AGN rather than high-mass stars. Such a population almost 
certainly exists. Both the brightest submillimetre-wave object detected by SIB
and the prototype of ultraluminous distant dusty galaxies, IRAS 
F\,10214+4724, contain AGN which must contribute to the heating of dust in 
these galaxies (Ivison et al.\ 1998b). In order to match the observed stellar density 
at the present epoch to that predicted in the models, but without modifying a 
standard Salpeter IMF, about 60\,per cent of the energy emitted by distant 
submillimetre-selected sources must be generated by AGN. Such a large fraction 
would require the lower limits to the dust temperatures derived in Section\,4 to 
be increased by a factor of about 20\,per cent, but this is not unreasonable. 
Sanders \& Mirabel (1996) report that about a third of ultraluminous galaxies 
detected by {\it IRAS} appear to contain an AGN. 

The relative importance of these alternative explanations cannot be determined
at present. Optical follow-up observations of submillimetre-selected sources 
will help to discriminate between the high-mass IMF and AGN models. The 
application of more advanced chemical enrichment models will also provide 
useful information when follow-up observations are available 
(Pei \& Fall 1995; Eales \& Edmunds 1996, 1997). 

\subsubsection{Generation of metals}

The mass of metals generated in any star-formation history can be calculated 
using equations (9) and (10). If divided by the solar abundance, assuming that 
the sun formed out of representative gas at $z=0.38$, the abundances of heavy
elements are predicted to evolve as shown in Fig.\,10(b) for the models listed in 
Table\,1. The data on metal abundances in damped Lyman-$\alpha$
systems determined by Pettini et al.\ (1997) are also plotted. As they noted, 
the fits to the optically derived counts and the Peak-G 
model are in full agreement with the data. By contrast, metals appear to be
overproduced by a  factor of about 5 at redshifts of 2 and 3 in the 
submillimetre-derived histories. However, in these histories the rate of
enrichment shown by the curves in Fig.\,10(b) is broadly consistent with the 
slope interpolated between the three highest redshift data points. 

Is this difference in the rate of production of metals significant? Clearly, modifying
the low-mass cut-off of the IMF as discussed above will not affect the metal 
enrichment rate. However, including a more substantial population of 
AGN-powered obscured sources at high redshifts would delay metal 
enrichment as compared with a model in which the sources were powered by 
star formation alone. If AGN outnumbered starbursts by a ratio of 4:1, then the 
enrichment curves in the submillimetre-selected models in Fig.\,10(b) would fall
into line with the data. Note that this fraction of AGN is greater than that
required to explain the discrepancy between the observed and predicted stellar 
density in Section\,5.3.1. 

Alternative arguments could also be used to explain at least some of the excess 
metal density predicted in the submillimetre-selected models. The observations 
of damped Lyman-$\alpha$ systems, from which the points in Fig.\,10(b) are 
derived, may sample the outer regions of disk systems (Ferguson, Gallagher \& 
Wyse 1998), rather than the highly-enriched cores, and so the metallicity of these 
systems may be systematically undersampled. In addition, quasars whose line of 
sight is intersected by the most metal-enriched systems are more likely to be 
obscured and thus less likely to be included in optically selected samples 
(Pei \& Fall 1995). If this were true, then the data in Fig.\,10(b) would tend to be 
obtained in directions with a systematically low metal density, again 
under-representing the true smoothed fraction of metals. The metal abundance
in intracluster gas observed by the {\it ASCA} X-ray observatory is typically 
about a third of the solar abundance (Gibson, Loewenstein \& Mushotzky 1998), a 
value which lies above the points in Fig.\,10(b), perhaps indicating that the global 
metal abundance at redshifts of 2 to 3 lies closer to the values predicted by 
the submillimetre-derived models. 

\subsubsection{Enrichment discussion}

The star formation histories in dusty galaxies that account for all of the 
far-infrared and submillimetre-wave data appear to be inconsistent with 
observations of $\Omega_*(0)$ and high-redshift metal enrichment. The results 
depend on the assumed IMF in dusty galaxies, and the optically derived data 
with which they are compared could well be subject to systematic biases. These 
biases would tend to lower the measured enrichment at high redshifts, and 
lead to a global value of enrichment in better agreement with the predictions of
the submillimetre-derived models, which we expect to be unaffected by 
either systematic biases or local effects. 

If the optical data is taken at face value, then either an IMF biased to high-mass 
star formation or a significant fraction, perhaps up to 60\,per cent, of distant 
ultraluminous submillimetre-wave sources must be powered by AGN in order to 
reconcile the submillimetre-wave and optical data. If the fraction of AGN at high 
redshifts reached about 60\,per cent, then the extinction-corrected star-formation 
history discussed by Pettini et al.\ (1998a,b) would account for most of the 
star-formation activity that has taken place in the Universe. 


More detailed follow-up observations of submillimetre-selected sources and 
detailed stellar and chemical evolution studies will be required in order to 
disentangle these effects. In particular the fraction of AGN uncovered in 
follow-up observations of submillimetre-wave surveys will provide a 
powerful test of these ideas (Ivison et al.\ 1998b). 

\subsection{Contamination by AGN}

Several circumstantial arguments can be used to illuminate the fraction of 
the SCUBA galaxies in which the dust is heated by AGN as compared with
star formation using existing data. 

The proportion of AGN that are missing from optically selected catalogues due 
to obscuration by dust has been discussed by Fall \& Pei (1993). The absorption 
of the optical emission from AGN by intervening dust is also discussed by 
Drinkwater et al. (1996). Fall \& Pei's models suggest that the true surface 
density of quasars with luminosities greater than of order $10^{12}$\,L$_\odot$ 
at redshifts in the range 2 to 3 -- sufficiently luminous to be detected by 
SCUBA -- should be of order 50\,deg$^{-2}$, after correcting for the effects of 
dust extinction. For comparison the surface density of SCUBA galaxies is 
about 2000\,deg$^{-2}$. 

Based on a radio selection technique, insensitive to the effects of dust 
absorption, Webster et al.\ (1995) estimate that 80\,per cent of radio-loud 
quasars are missing from optically selected samples. If the same fraction 
of radio-quiet quasars is absent, then the surface density of 
obscured quasars that are sufficiently bright to be detected by SCUBA 
should be of order 100\,deg$^{-2}$.  The results of deep X-ray surveys are 
also insensitive to the effects of obscuration by dust, and resolve a significant 
fraction of the X-ray background radiation. In order to account for the 
background spectrum, a population of dust obscured AGN, which are not 
detected in optical surveys, must be included in models (Almaini et al.\ 1997, 
1998; Almaini, Lawrence \& Boyle 1999). The surface density of faint X-ray 
sources, which are candidates for this population, is about 
$230\pm40$\,deg$^{-2}$ (Jones et al. 1997). 

These observations all tend to suggest that the surface density of luminous  
dust obscured AGN might be typically up to a few tenths of the surface density 
of SCUBA galaxies, and thus that star-formation activity is responsible for 
the greater fraction of the submillimetre-wave flux density emitted by the
population of SCUBA galaxies. This proportion is consistent with the fraction 
of compact sources identified with SCUBA galaxies by Smail et al. (1998). 

Only one SCUBA-selected source, SMM\,J02399$-$0136, has been subject to 
detailed follow-up observations (Ivison et al. 1998). While SMM\,J02399$-$0136 
certainly contains an AGN, a significant fraction of its luminosity is thought to 
be produced by massive star formation. The recent detection of redshifted 
CO(3$\rightarrow$2) emission from SMM\,J02399$-$0136 using the Owens Valley 
Millimeter Array (Frayer et al. 1998) supports the idea that approximately equal 
amounts of its luminosity are contributed by star formation activity and an AGN. 

If the SCUBA sources are powered predominantly by star formation activity, then 
a global gas consumption rate of order 0.5\,M$_\odot$\,yr$^{-1}$\,Mpc$^{-3}$ is
required at redshifts of about 3. This rate can be compared with the predictions of 
the self-consistent chemical evolution models presented by Pei \& Fall (1995). 
The infall model of Pei \& Fall (1995) predicts a very similar consumption rate, 
and so there is no practical limit to the fraction of the luminosity of SCUBA 
sources that can be generated by star formation rather than AGN. Detailed 
multiwaveband follow-up observations will be required to determine the 
relative importance of AGN and star formation for heating dust in SCUBA 
galaxies. At present it seems likely that up to about a third of the emission from 
these sources is powered by AGN, with the remainder due to star formation. 
Interestingly, this is the fraction of the count of SCUBA galaxies predicted on the 
basis of the X-ray background and luminosity function by Almaini et al.\ (1999). 

\section{The consequences for future observations}

The submillimetre-derived star formation histories shown in Fig.\,9 correspond 
to models of galaxy evolution that account for the low-redshift evolution of 
{\it IRAS} galaxies, the observed submillimetre-wave background radiation 
intensity and the 850-$\mu$m counts, and predict counts which are
in agreement with  the limits determined at wavelengths of 2.8\,mm and 
450\,$\mu$m. The predicted background radiation intensities, source counts and 
redshift distributions at a 
wide range of wavelengths are derived in these models and the results are used 
to discuss the future of submillimetre-wave cosmology. The most promising 
observations with which to obtain more accurate determinations of the history of 
obscured star formation are considered. 

\begin{figure}
\begin{center}
\epsfig{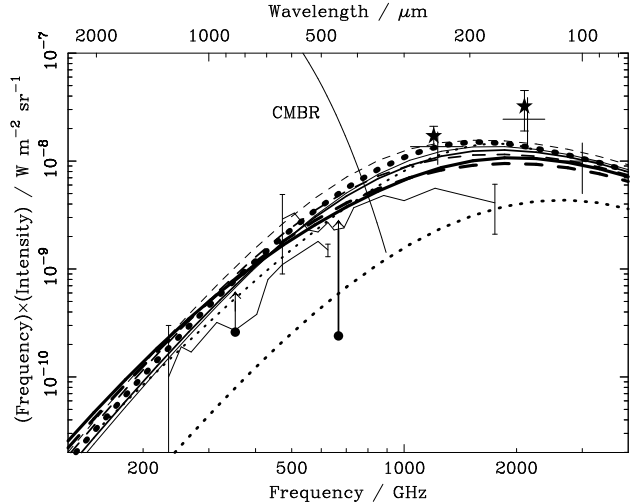} 
\end{center}
\caption{Predicted background radiation intensities in the eight models of galaxy 
formation listed in Table\,1 that are fitted to submillimetre-wave data. The 
background data is plotted in the format of Fig.\,2. 
}
\end{figure}

\subsection{Background radiation}

Predictions for the intensity of background radiation in the
millimetre/submillimetre and far-infrared wavebands derived in the models listed 
in Table\,1 are shown in Fig.\,11. Similar spectra are predicted by all the
models, regardless of their very different star formation histories. The ensemble
of models included in Fig.\,11 predicts background intensities of 
$(8.3\pm1.2)\times10^{-10}$ and 
$(4.6\pm0.8)\times10^{-9}$\,W\,m$^{-2}$\,sr$^{-1}$ at wavelengths of 
850 and 450\,$\mu$m respectively. For comparison, the corresponding values 
reported by Puget et al. (1996) and Fixsen et al. (1998) are 
$(2.7\pm2.0)\times10^{-10}$, $(2.3\pm2.0)\times10^{-9}$ and 
$(5.3\pm1.6)\times10^{-10}$, 
$(3.3\pm1.0)\times10^{-9}$\,W\,m$^{-2}$\,sr$^{-1}$ respectively. The form of
the background spectrum is therefore not a good quantity with which to
discriminate between different models of the star-formation history. 

\subsection{Source counts}

The source counts predicted at wavelengths of 2.8\,mm, 1.3\,mm, 850\,$\mu$m, 
450\,$\mu$m, 175\,$\mu$m, 60\,$\mu$m and 15\,$\mu$m are shown in Fig.\,12. 
The observed counts derived by Wilner \& Wright (1997), SIB, 
Holland et al.\ (1998b), Kawara et al.\ (1997), the authors listed in Fig.\,5, and 
Oliver et al.\ (1997) are also plotted. The predicted abundance of lensed
submillimetre-wave galaxies (Blain 1996) is discussed elsewhere (Blain 1998c). 

The 15-$\mu$m counts are included for completeness only. The models used
here have little predictive power at wavelengths shorter than about 40\,$\mu$m, 
and correspondingly the 15-$\mu$m counts have little predictive power 
for determining the star formation history at high redshifts. The median redshift 
of the sources detected by {\it ISO} at 15\,$\mu$m is expected to be less than 0.5 
(see Fig.\,13a). The 15-$\mu$m counts are uniformly underpredicted by a 
factor of about 3, about the size of the observation uncertainty. The 
predicted count curves can be made to traverse the 15-$\mu$m count error bar 
shown in Fig.\,12(b), by sweeping the value of the high-frequency power-law 
index in the spectral energy distribution model (see Section\,3.2) through the 
range $-1.4$ to $-2.2$. The value of this power-law index has almost no effect 
on either the source counts or background radiation spectra expected at longer 
wavelengths. 
  
\begin{figure*}
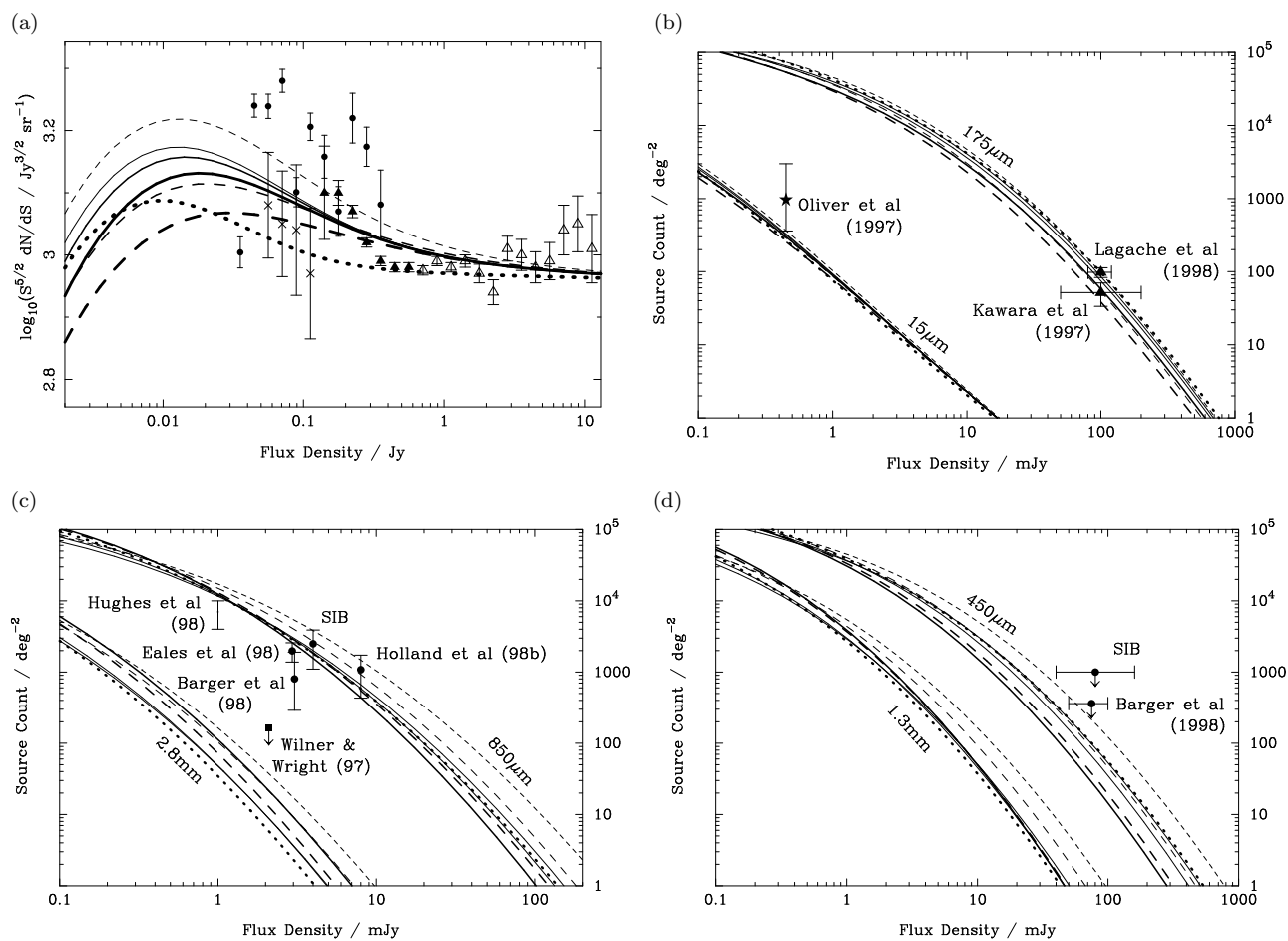

\begin{minipage}{170mm}
(a) \hskip 81mm (b)
\begin{center}
\vskip-4mm
\epsfig{file=fig12a.ps, width=5.6cm, angle=-90} \hskip 5mm 
\epsfig{file=fig12b.ps, width=5.75cm, angle=-90}
\end{center}
(c) \hskip 81mm (d)
\begin{center}
\vskip -4mm
\epsfig{file=fig12c.ps, width=5.6cm, angle=-90} \hskip 4mm
\epsfig{file=fig12d.ps, width=5.6cm, angle=-90}
\end{center}
\caption{Counts predicted by the models listed in Table\,1, as compared with 
available data. The 60-$\mu$m counts are shown in (a), those at 15 and 
175\,$\mu$m are shown in (b), those at 850\,$\mu$m and 2.8\,mm are shown in
(c), and those at 450\,$\mu$m and 1.3\,mm are shown in (d). The
references to the data in (a) are given in the caption of Fig.\,5; in (b), (c) and (d) 
they are written adjacent to the data points.}
\end{minipage}
\end{figure*}

\begin{table}
\caption{The redshifts below which 10, 50 and 90 per cent of sources detected 
in surveys at reasonable flux density limits at a range of wavelengths are 
expected to lie, $\bar z_{10}$, $\bar z_{50}$ and $\bar z_{90}$ respectively. The 
values are obtained from the spread of the redshift distribution curves in Fig.\,13. 
}
{\vskip 0.75mm}
\hrule{\vskip 1.2mm}
\begin{tabular}{ p{1.4cm} p{1.4cm} p{1.3cm} p{1.3cm} p{1.2cm}}
Wavelen-& Flux limit & $\bar z_{10}$ & $\bar z_{50}$ & $\bar z_{90}$ \\
gth / $\mu$m & / mJy & & & \\
\noalign{
{\vskip 1.2mm}
\hrule
{\vskip 2.7mm}
}
15 & 0.4 & 0.07--0.13 & 0.32--0.53 & 1.2--1.9\\ 
60 & 10 & 0.07--0.13 & 0.32--0.53 & 1.2--1.9\\
175 & 100 & 0.17--1.0
& 1.1--1.8 & 1.8--2.7\\
450 & 50 & 1.3--2.0 & 2.2--3.1
& 2.8--4.5\\
850 & 2 & 1.7--2.2
& 2.4--4.4
& 3.8--8.2\\
1300 & 2 & 1.7--2.7
& 2.5--5.0
& 3.8--9.2\\
2800 & 2 & 1.8--3.6
& 2.5--7.8
& 4.0--9.8\\
\noalign{\vskip 1.2mm}
\end{tabular}
\hrule
\end{table}

\begin{figure*}
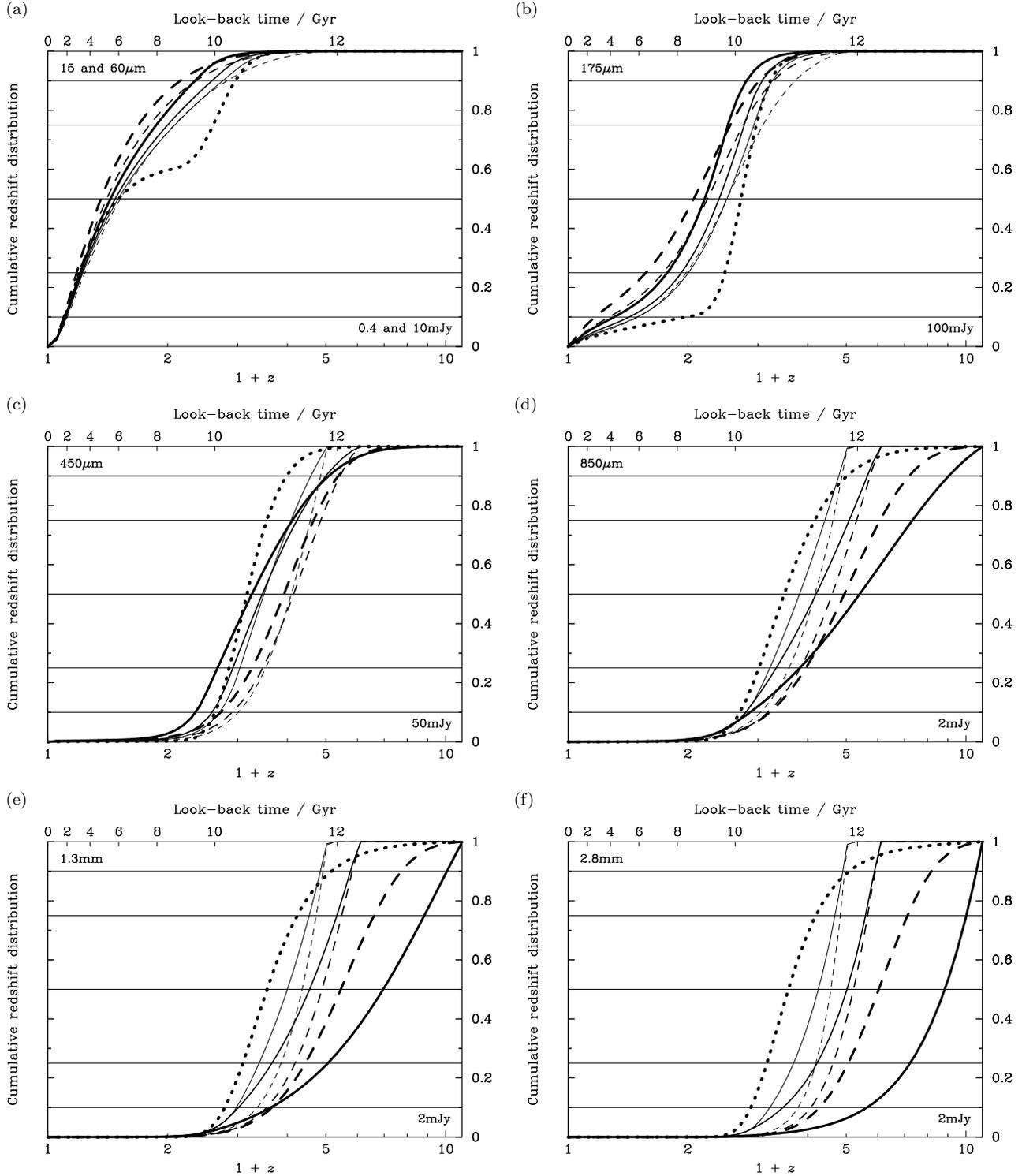

\begin{minipage}{170mm}
(a) \hskip 81mm (b)
\begin{center}
\vskip -6mm
\epsfig{file=fig13a.ps, width=6.25cm, angle=-90} \hskip 5mm
\epsfig{file=fig13b.ps, width=6.25cm, angle=-90}
\end{center}
(c) \hskip 81mm (d)
\begin{center}
\vskip -6mm
\epsfig{file=fig13c.ps, width=6.25cm, angle=-90} \hskip 5mm
\epsfig{file=fig13d.ps, width=6.25cm, angle=-90}
\end{center}
(e) \hskip 81mm (f)
\begin{center}
\vskip -6mm
\epsfig{file=fig13e.ps, width=6.25cm, angle=-90} \hskip 5mm
\epsfig{file=fig13f.ps, width=6.25cm, angle=-90}
\end{center}
\caption{Predicted cumulative redshift distributions for submillimetre-selected 
sources detected at the wavelengths and flux density limits listed in Table\,2. 
The styles and widths of the plotted curves correspond to the key in Table\,1. 
The Peak-G model is not plotted. The redshift distributions at 
15\,$\mu$m/0.4\,mJy and 60\,$\mu$m/10\,mJy are expected to be very similar. A 
range of surveys with different flux density limits at wavelengths of 450 and 
850\,$\mu$m will soon be completed. The redshift distributions predicted at a 
flux density limit of 10\,mJy were also calculated at both these wavelengths, and
found to be very similar to those shown in (c) and (d). The results of such 
surveys can therefore be compared directly with the predictions made here.
Based on the colours of the optical counterparts to the faint SCUBA-selected 
sources from the extended cluster survey described by Smail et al.\ (1998), 
greater than $75\pm20$\,per cent and $50\pm15$\,per cent of the counterparts 
have redshifts less than 5.5 and 4.5 respectively.  
}
\end{minipage}
\end{figure*}

The spread in the counts predicted across the range of well-fitting models listed 
in Table\,1 offers excellent prospects for discriminating between the models 
using submillimetre-wave counts determined at a range of flux densities and 
observing wavelengths. A range of existing and future facilities will be useful 
for determining counts; {\it Planck Surveyor} (Bersanelli et al.\ 1996), 
large ground-based millimetre/submillimetre-wave interferometer arrays (Brown 
et al.\ 1996; Downes et al.\ 1996), advanced-technology single-antenna
ground-based telescopes (Schloerb 1997; Stark et al.\ 1998) and {\it FIRST}
(Pibratt 1997). 

\subsection{Redshift distributions}

The redshift distributions of submillimetre-selected sources at, or just below, the 
flux density limits of current surveys are shown in Fig.\,13. The sensitivity 
thresholds and the redshift ranges within which the redshift distributions of
source in these surveys are expected to be 10, 50 and 90\,per cent complete are 
listed in Table\,2. Based on the values listed in Table\,2, the median redshift of 
SIB's submillimetre-selected sample should be in the range 2.4 to 4.4. 
This is in broad agreement with constraints obtained from broadband
photometry of the optical counterparts by Smail et al.\ (1998),
as well as the first spectroscopic identifications (Ivison et al.\ 1998b).
  
The crucial result revealed in Table\,2 is that most of the sources detected in 
submillimetre-wave surveys are likely to lie at redshifts well below 10, and so a 
large proportion will be accessible to multi-waveband study using 8-m-class 
telescopes. The advantage of a cluster lens survey, of course, is that in addition 
to the rich archival data available, the sources are typically magnified by a factor 
of a few, both in the submillimetre and the optical waveband. Not only does this 
make it easier to detect these faint sources in the  submillimetre waveband, but it 
also facilitates spectroscopic follow-up observations as compared with a 
blank-field survey. Blank-field submillimetre-wave surveys have wisely 
exploited the most carefully surveyed areas of the field sky -- the {\it Hubble 
Deep Field}, CFRS fields and Hawaii Deep Survey fields (Hughes et al.\ 1998; 
Eales et al.\ 1998; Barger et al.\ 1998).

If confirmed by spectroscopic observations, the relatively modest median redshift 
predicted for submillimetre-selected galaxies would further suggest that they 
comprise a distinct population of strongly obscured sources that have evaded 
detection in optical surveys on the grounds of their extreme obscuration rather 
than their extreme distance.

\subsection{Consequences for earlier results}

In earlier papers we discussed the first results of our survey (SIB) and the 
consequences of our detection of a population of distant dusty galaxies for 
source confusion in existing and future observations (BIS; Blain et al. 1998b). 
The first results of follow-up observations in other wavebands have also been 
published (Ivison et al.\ 1998b). The results presented here reinforce the 
conclusions presented in these earlier papers, demonstrating that simple 
well-constrained and self-consistent families of models of the evolution of 
distant obscured galaxies and quasars can account for all the available data in 
the millimetre, submillimetre and far-infrared wavebands. A more complete 
description of the consequences for future surveys lies outside the scope of 
this paper: see Blain et al.\ (1998b) and Blain (1998c).

\section{Conclusions}

We have presented a family of self-consistent models of the evolution of 
dusty galaxies that account for all the available data in the submillimetre and 
far-infrared wavebands. 

\begin{enumerate}

\item The results suggest that the luminosity density emitted in the far-infrared
waveband at $z>1$ evolves in a very different way to that inferred from 
observations in the optical waveband. By integrating our new luminosity 
densities over all epochs, we find that the total optically-derived volume 
emissivity (Madau et al. 1996) underpredicts the true value by a factor of about 4. 
Furthermore, we cannot reproduce the results of submillimetre-wave
observations using a volume emissivity that evolves in the same way as the 
optically derived form. The peak of the comoving submillimetre-derived 
luminosity density also occurs at a considerably higher redshift than that 
suggested by optical observations, $z\sim 2$--5. 

\item If our submillimetre-derived star-formation histories are associated with a 
Salpeter IMF with a lower mass limit of 0.07\,M$_\odot$, then the number of 
low-mass stars at the present epoch and the apparent metallicity at 
$z \simeq 2$--3 are overproduced by factors of up to about 5. 

\item In order to account for the discrepancy in
the abundance of low-mass stars, then either the IMF in distant dusty galaxies 
must have a low-mass cut-off of about 0.7\,M$_\odot$, or up to 60\,per 
cent of the energy emitted by distant dusty galaxies must be generated by 
gravitational accretion in AGN. 

\item In order to understand the discrepancy between the metal enrichment 
at moderate redshifts, either gravitational accretion in AGN must be responsible
for a large fraction of the energy emitted by distant dusty galaxies or 
the current optical observations are subject to considerable selection effects
that lead to the underprediction of the metal content at $z\simeq 2$--3. 

\item The current evidence suggests that the population of distant luminous
submillimetre-selected sources is missing from surveys made in the optical 
waveband because of their very considerable dust extinction, rather than because
they lie at extremely great distances.

\end{enumerate} 

Using the results presented in this paper, we can determine the most appropriate
strategy for making further progress in understanding the nature of distant dusty
galaxies and their relationship with optically-selected distant galaxies: 

\begin{enumerate} 

\item In order to learn more about the evolution of dusty galaxies, measurements 
of the submillimetre-wave counts at wavelengths longer than 200\,$\mu$m will 
allow the parameters that define the families of well-fitting models to be 
determined more accurately. However, it will still not be easy to discriminate 
between different models within the same family using more accurate counts. 

\item In order to address the relationship between sub-millimetre-selected 
samples and optically selected samples, follow-up optical and near-infrared 
imaging and spectroscopy of the submillimetre-selected sources will be required. 
On the basis of our predicted redshift distributions for the submillimetre-selected 
sources, it seems likely that a significant fraction of the sources will be 
detectable using the best existing optical and near-infrared instruments.

\item The redshift distribution of submillimetre-selected sources will allow us
to discriminate between different members of the same well-fitting family of 
galaxy formation models that can account for all the submillimetre-wave source 
count and background radiation data. The determination of a substantially 
complete redshift distribution is the key goal of follow-up observations during 
the course of the next year. 
 
\item The optical and near-infrared spectra obtained to derive such a redshift 
distribution will be crucial for determining the relative 
importance of star-forming galaxies and AGN in submillimetre-selected samples, 
and therefore to investigate the apparent discrepancy between the rate of metal 
enrichment inferred from optical and submillimetre-wave observations. 

\end{enumerate}

We now have a consistent picture of the evolution of the submillimetre-wave 
Universe; however, its relationship to our understanding derived from optical 
observations remains unclear. Some part of the discrepancy will be
explained by the accurate quantification of known selection biases in the optical 
observations. The crucial information required to investigate the history of star 
formation in more detail will be provided by larger submillimetre-wave surveys, 
and, most importantly, by optical and near-infrared follow-up observations of 
submillimetre-selected sources. If these observations confirm that most
submillimetre-selected sources are distant and show no signs of AGN activity, 
then there is a strong contradiction to investigate. 

\section*{Acknowledgements}
We thank Malcolm Longair, the referee Priya Natarajan, Max Pettini, Richard 
Saunders, Roberto Terlevich and Neil Trentham for helpful comments on the 
manuscript. We thank the commissioning team of SCUBA, and Ian Robson for his 
continuing support and enthusiasm. AWB, IRS and RJI are supported by PPARC.


\begin{thebibliography}{} 

\bibitem[\protect\citename{blah}%
]{Al1}
Almaini O., Shanks T., Griffiths R.\,E., Boyle B.\,J., Roche N., Georgantopoulos I., 
Stewart G.\,C., 1997, MNRAS, 291, 372

\bibitem[\protect\citename{blah}%
]{Al2}
Almaini O., Shanks T., Gunn K.\,F., Boyle B.\,J.,
Georgantopoulos I., Griffiths R.\,E., Stewart G.\,C., Blair A.\,J.,
1998, Astron. Nachr., 319, 55 (astro-ph/9711150)

\bibitem[\protect\citename{blah}%
]{Al3}
Almaini O., Lawrence A., Boyle B.\,J., 1999, MNRAS, submitted

\bibitem[\protect\citename{blah}%
]{Barger}
Barger A.\,J., Cowie L.\,L., Sanders D.\,B., Fulton E., Taniguchi Y., 
Sato Y., Kawara K., Okuda H., 1998, Nat,  394, 248

\bibitem[\protect\citename{bl}%
]{baugh}
Baugh C.\,M., Cole S.\,M., Frenk C.\,S., Lacey C.G., 1998, ApJ, 498, 504

\bibitem[\protect\citename{bl}%
]{Ben}
Benford D.\,J., Cox P., Omont A., Phillipps T.\,G., 1998, BAAS, 30, 783

\bibitem[\protect\citename{bl}%
]{PLANCK}
Bersanelli M. et al., 1996, COBRAS/SAMBA. SCI(96)3, ESA, Paris

\bibitem[\protect\citename{bl}%
]{Bertin}
Bertin E., Dennefeld M., Moshir M., 1997, A\&A, 323, 685

\bibitem[\protect\citename{bl}%
]{Biller}
Biller S.\,D. et al., 1998, PRL, 80, 2992 (astro-ph/9802234)

\bibitem[\protect\citename{bl}%
]{BL96B}
Blain A.\,W., 1996, MNRAS, 283, 1340

\bibitem[\protect\citename{bl}%
]{BL97B}
Blain A.\,W., 1997, MNRAS, 290, 553

\bibitem[\protect\citename{bl}%
]{BL98a}
Blain A.\,W., 1998a, MNRAS, 295, 92 (astro-ph/9710160)

\bibitem[\protect\citename{bl}%
]{BL98b}
Blain A.\,W., 1998b, MNRAS, 297, 511 (astro-ph/9801098)


\bibitem[\protect\citename{bl}%
]{BL98c}
Blain A.\,W., 1998c, in Colombi S., Mellier S. eds, Wide-field surveys
in cosmology, Proc. XIV IAP Conference, Editions Fronti\`eres, Gif-Sur-Yvette, 
in press (astro-ph/9806369)

\bibitem[\protect\citename{bl}%
]{BL93A}
Blain A.\,W., Longair M.\,S., 1993a, MNRAS, 264, 509

\bibitem[\protect\citename{bl}%
]{BL93B}
Blain A.\,W., Longair M.\,S., 1993b, MNRAS, 265, L21

\bibitem[\protect\citename{bl}%
]{BL96}
Blain A.\,W., Longair M.\,S., 1996, MNRAS, 279, 847

\bibitem[\protect\citename{bl}%
]{BIS}
Blain A.\,W., Ivison R.\,J., Smail I., 1998a, MNRAS, 296, L29 (astro-ph/9710003; BIS)

\bibitem[\protect\citename{bl}%
]{BISKParis}
Blain A.\,W., Ivison R.\,J., Smail I., Kneib J.-P., 1998b, in Colombi S., 
Mellier Y. eds, 
Wide-field surveys in cosmology, Proc. XIV IAP Conference, Editions Fronti\`eres, 
Gif-Sur-Yvette, in press (astro-ph/9806063) 

\bibitem[\protect\citename{bl}%
]{BKIS}
Blain A.\,W., Kneib J.-P., Ivison R.\,J., Smail I., 1998c, ApJ, submitted 

\bibitem[\protect\citename{bl}%
]{BT}
Boyle B.\,J., Terlevich R., 1998, MNRAS, 293, L49


\bibitem[\protect\citename{bl}%
]{BROWN}
Brown R.\,L., 1996, in Bremer M.\,N., van der Werf P., R\"ottgering~H.\,J.\,A., Carilli
C.\,R. eds., Cold Gas at High Redshift, Kluwer, Dordrecht, p.\,411

\bibitem[\protect\citename{bl}%
]{Burigana}
Burigana C., Danese L., De Zotti G., Franceschini A., Mazzei P., Toffolatti L., 1997, 
MNRAS, 287, L17

\bibitem[\protect\citename{bl}%
]{BTyt}
Burles S., Tytler D., 1998, ApJ, 499, 699 (astro-ph/9712108)

\bibitem[\protect\citename{blah}%
]{CALZETTI}
Calzetti D., Bohlin R.\,C., Kinney A.\,L., Storchi-Bergmann T., Heckmann T.\,M.,
1995, ApJ, 443, 136

\bibitem[\protect\citename{blah}%
]{Cim}
Cimatti A., Andreani P., R\"ottgering H.\,J.\,A., Tilanus R., 1998, Nat, 392, 895

\bibitem[\protect\citename{blah}%
]{CLOSE}
Close L.\,M., Hall P.\,B., Liu C.\,T., Hege E.\,K., 1995, ApJ, 452, L9

\bibitem[\protect\citename{blah}%
]{CAFNZ}
Cole S.\,M., Arag\'on-Salamanca A., Frenk C.\,S., Navarro J.\,F.,
Zepf S.\,E., 1994, MNRAS, 271, 781.

\bibitem[\protect\citename{blah}%
]{Con}
Connolly A.\,J., Szalay A.\,S., Dickinson M., SubbaRao M.\,U., Brunner R.\,J., 
1997, ApJ, 486, L11 (astro-ph/9706255)

\bibitem[\protect\citename{blah}%
]{Cram}
Cram L.\,E., 1998, ApJ, 506, L85 (astro-ph/9808228) 
\
\bibitem[\protect\citename{bl}%
]{Dey}
Dey A., Graham J.\,R., Ivison R.\,J., Smail I., 1998, ApJ, submitted

\bibitem[\protect\citename{bl}%
]{DOWNES}
Downes D., 1996, in Shaver P. ed., Science with Large 
Millimetre Arrays. Springer, Berlin, p.\,16

\bibitem[\protect\citename{bl}%
]{Drink}
Drinkwater M.\,J., Webster R.\,L., Francis P.\,J., Wilklind T., Combes F., 1996, 
Proc. Astron. Soc. Aust., 13, 183

\bibitem[\protect\citename{bl}%
]{Dun}
Dunlop J.\,S., 1998, in Bremer M.\,N., Jackson N., Perez-Fournon I.\ eds, 
Observational Cosmology with the New Radio Surveys. Kluwer, Dordrecht, p.\,157 
(astro-ph/9704294)

\bibitem[\protect\citename{bl}%
]{DP}
Dunlop J.\,S., Peacock J.\,A., 1990, MNRAS, 247, 19

\bibitem[\protect\citename{bl}%
]{DS}
Dwek E., Slavin J., 1994, ApJ, 436, 696

\bibitem[\protect\citename{bl}%
]{Dwek98}
Dwek E. et al., 1998, ApJ, in press (astro-ph/9806129) 

\bibitem[\protect\citename{bl}%
]{EE96}
Eales S.\,A., Edmunds M.\,G., 1996, MNRAS, 280, 1167

\bibitem[\protect\citename{bl}%
]{EE97}
Eales S.\,A., Edmunds M.\,G., 1997, MNRAS, 286, 732

\bibitem[\protect\citename{bl}%
]{Eetal}
Eales S.\,A., Lilly S.\,J., Gear W.\,K., Dunne L., Bond J.\,R., Hammer F., 
Le F\`evre O., Crampton D., 1998, ApJL, submitted (astro-ph/9808040) 

\bibitem[\protect\citename{bl}%
]{FP}
Fall S.\,M., Pei Y.\,C., 1993, ApJ, 402, 479

\bibitem[\protect\citename{bl}%
]{FGW98}
Ferguson A.\,M.\,N., Gallagher J., Wyse R.\,F.\,G., 1998, AJ, 116, 673
(astro-ph/9805166)

\bibitem[\protect\citename{blah}%
]{FIXinst}
Fixsen D.\,J. et al.\ 1994, ApJ, 420, 457

\bibitem[\protect\citename{blah}%
]{FIX}
Fixsen D.\,J., Cheng E.\,S., Gales J.\,M., Mather J.\,C., Shafer R.\,A., Wright E.\,L.,
1996, ApJ, 473, 576

\bibitem[\protect\citename{blah}%
]{FIXNEW}
Fixsen D.\,J., Dwek E., Mather J.\,C., Bennett C.\,L., Shafer R.\,A., 1998, ApJ, 
in press (astro-ph/9803021)

\bibitem[\protect\citename{blah}%
]{FB}
Frayer D.\,T., Brown R.\,L., 1997, ApJS, 113, 221

\bibitem[\protect\citename{blah}%
]{Fetal}
Frayer D.\,T., Ivison R.\,J., Scoville N.\,Z., Yun M.\,S., Evans A.\,S., Smail I., 
Blain A.\,W., Kneib J.-P., 1998, ApJ, 506, L7 (astro-ph/9808109) 

\bibitem[\protect\citename{blah}%
]{Gal}
Gallego J., Zamorano J., Arag\'on-Salamanca A., Rego M., 1996, ApJ, 459, L43

\bibitem[\protect\citename{blah}%
]{GLM}
Gibson B.\,K., Loewenstein M., Mushotzky R.\,F., 1998, MNRAS, 290, 623

\bibitem[\protect\citename{blah}%
]{GnO}
Gnedin N.\,Y., Ostriker J.\,P., 1992, ApJ, 400, 1

\bibitem[\protect\citename{blah}%
]{Getal}
Goldschmidt P. et al., 1997, MNRAS, 289, 465

\bibitem[\protect\citename{blah}%
]{GD}
Graham J.\,R., Dey A., 1996, ApJ, 471, 720

\bibitem[\protect\citename{blah}%
]{G}
Gregorich D.\,T., Neugebauer G., Soifer B.\,T., Gunn J.\,E., Hertler T.\,L., 1995, AJ,
110, 259

\bibitem[\protect\citename{blah}%
]{Gron}
Gronwall C., 1998, in Thuan T.\,X. et al. eds, 
Dwarf galaxies and cosmology. Proc.
XVIII Moriond meeting, Editions Fronti\`eres, Gif-sur-Yvette, in press 
(astro-ph/9806240)

\bibitem[\protect\citename{blah}%
]{Guiderdoni}
Guiderdoni B., Bouchet F.\,R., Puget J.-L., Lagache G., Hivon E., 1997, 
Nat, 390, 257

\bibitem[\protect\citename{blah}%
]{HH}
Hacking P.\,B., Houck J.\,R., 1987, ApJS, 63, 311

\bibitem[\protect\citename{blah}%
]{HF}
Hammer F., Flores H., 1998, in Thuan T.\,X. et al. eds, Dwarf galaxies 
and cosmology. 
Proc. XVIII Moriond meeting, Editions Fronti\`eres, Gif-sur-Yvette, in press 
(astro-ph/9806184)

\bibitem[\protect\citename{blah}%
]{H96}
Hauser M.\,G., 1996, in Kafatos M., Kondo Y. eds, Examining the Big Bang and 
Diffuse Background Radiations. Proc. IAU 168, Kluwer, Dordrecht, p.\,99 

\bibitem[\protect\citename{blah}%
]{H96}
Hauser M.\,G. et al., 1998, ApJ, in press (astro-ph/9806167)

\bibitem[\protect\citename{bl}%
]{HFC}
Hewett P.\,C., Foltz C.\,B., Chaffee F.\,H., 1993, ApJ, 406, L43

\bibitem[\protect\citename{bl}%
]{Holl}
Holland W.\,S. et al., 1999, MNRAS, in press (astro-ph/9809122) 

\bibitem[\protect\citename{bl}%
]{Holl2}
Holland W.\,S. et al., 1998, Nat, 392, 788 

\bibitem[\protect\citename{bl}%
]{HughesCG}
Hughes D.\,H., 1996, in Bremer M.\,N., van der Werf P.\,P., R\"ottgering H.\,J.\,A., 
Carilli C.\,L. eds, Cold Gas at High Redshift. Kluwer Dordrecht, p.\,311

\bibitem[\protect\citename{bl}%
]{Hughes}
Hughes D.\,H. et al., 1998, Nat, 394, 241 (astro-ph/9806297) 

\bibitem[\protect\citename{bl}%
]{I97}
Ivison R.\,J., Archibald E.\,N., Dey A., Graham J.\,R., 1997, in Wilson A. ed., 
The Far-Infrared and Submillimetre Universe. ESA publications, Noordwijk, p.\,281

\bibitem[\protect\citename{bl}%
]{I}
Ivison R.\,J. et al., 1998a, ApJ, 494, 211 (astro-ph/9709124)

\bibitem[\protect\citename{bl}%
]{I+7}
Ivison R.\,J., Smail I., Le Borgne J.-F., Blain A.\,W., Kneib J.-P., B\'ezecourt J., 
Kerr T.\,H., Davies J.\,K., 1998b, MNRAS, 298, 583 (astro-ph/9712161)

\bibitem[\protect\citename{bl}%
]{J}
Jones L.\,R. et al., 1997, MNRAS, 285, 547


\bibitem[\protect\citename{bl}%
]{Kawara}
Kawara K. et al., 1997, in Wilson A.\ ed.,\ The Far-Infared and Submillimetre 
Universe. ESA publications, Noordwijk, p.\,285

\bibitem[\protect\citename{bl}%
]{Kenn}
Kennicutt R.\,C., 1993, ApJ, 272, 54 

\bibitem[\protect\citename{bl}%
]{Lagache}
Lagache G. et al., 1998, in Colombi S., Mellier Y. eds, 
Wide-field surveys in cosmology, Proc. XIV IAP Conference, Editions Fronti\`eres, 
Gif-Sur-Yvette, in press

\bibitem[\protect\citename{bl}%
]{Larson}
Larson R.\,B., 1998, MNRAS, in press (astro-ph/9808145)

\bibitem[\protect\citename{bl}%
]{L}
Lilly S.\,J., Le F\`evre O., Hammer F., Crampton D., 1996, ApJ, 460, L1


\bibitem[\protect\citename{bl}%
]{Madau}
Madau P., Ferguson H.\,C., Dickinson M.\,E., Giavalisco M., Steidel~C.\,C., 
Fruchter A., 1996, MNRAS, 283, 1388

\bibitem[\protect\citename{bl}%
]{Madau2}
Madau P., Della Valle M., Panagia N., 1998, MNRAS, 297, L17

\bibitem[\protect\citename{bl}%
]{MB}
Mazzarella J.\,M., Balzano V.\,A., 1986, ApJS, 62, 751

\bibitem[\protect\citename{bl}%
]{ML} 
Mushotzky R.\,F., Loewenstein M., 1997, ApJ, 481, L63

\bibitem[\protect\citename{bl}%
]{ORS}
Oliver S.\,J., Rowan-Robinson M., Saunders W., 1992, MNRAS, 256, 15P
 
\bibitem[\protect\citename{bl}%
]{Oet al}
Oliver S.\,J. et al.\ 1997, MNRAS, 289, 471

\bibitem[\protect\citename{bl}%
]{PRR}
Pearson C., Rowan-Robinson M., 1996, MNRAS, 283, 174

\bibitem[\protect\citename{bl}%
]{PF}
Pei Y.\,C., Fall S.\,M., 1995, ApJ, 454, 69

\bibitem[\protect\citename{bl}%
]{Pettini2}
Pettini M., Smith L.\,J., King D.\,L., Hunstead R.\,W., 1997, ApJ, 486, 665

\bibitem[\protect\citename{bl}%
]{Pettini3}
Pettini M., Kellogg M., Steidel C.\,C., Dickinson M., Adelberger~K.\,L., 
Giavalisco M., 1998a, ApJ, in press (astro-ph/9806219)

\bibitem[\protect\citename{bl}%
]{Pettini}
Pettini M., Steidel C.\,C., Adelberger K.\,L., Kellogg M., Dickinson M., Giavalisco M.,
1998b, in Shull J.\,M., Woodward C.\,E., Thronson H.\,A.\ eds, Cosmic
Origins: evolution of galaxies, stars, planets and life. Astr. Soc. Pac., San 
Francisco, p. 67 (astro-ph/9708117)

\bibitem[\protect\citename{bl}%
]{PH}
Phillipps T.\,G., 1997, in Wilson A.\ ed.,\ The far-infrared and
submillimetre universe. ESA publications, Noordwijk, p.\,223

\bibitem[\protect\citename{bl}%
]{FIRST}
Pilbratt G., 1997, in Wilson A.\ ed.,\ 
The Far-Infrared and Submillimetre Universe. ESA publications, Noordwijk, p.\,7

\bibitem[\protect\citename{blah}%
]{PABB}
Puget J.-L., Abergel A., Bernard J.-P., Boulanger F., Burton~W.\,B., D\'esert
F.-X., Hartmann D., 1996, A\&A, 308, L5

\bibitem[\protect\citename{bl}%
]{Rieke}
Rieke G.\,H., Loken K., Rieke M.\,J., Tamblyn P., 1993, ApJ, 412, 99


\bibitem[\protect\citename{bl}%
]{RR90}
Rowan-Robinson M., Hughes J., Vedi K., Walker D.\,W., 1990, MNRAS, 246, 273

\bibitem[\protect\citename{bl}%
]{RR91}
Rowan-Robinson M. et al., 1991, Nat, 351, 719

\bibitem[\protect\citename{bl}%
]{RR97}
Rowan-Robinson M. et al., 1997, MNRAS, 289, 490

\bibitem[\protect\citename{bl}%
]{SM}
Sanders D.\,B., Mirabel I.\,F., 1996, ARA\&A, 34, 749

\bibitem[\protect\citename{bl}%
]{IRAS}
Saunders W., Rowan-Robinson M., Lawrence A., Efstathiou G., Kaiser N., 
Ellis R. S., Frenk C.\,S., 1990, MNRAS, 242, 318

\bibitem[\protect\citename{bl}%
]{SS}
Savage B.\,D., Sembach K.\,R., 1996, ARA\&A, 34, 279

\bibitem[\protect\citename{blah}%
]{SFD}
Schlegel D.\,J., Finkbeiner D.\,P., Davis M., 1998, ApJ, 500, 525
(astro-ph/9710327)

\bibitem[\protect\citename{blah}%
]{LMT}
Schloerb F.\,P., 1997, in Latter W.\,B., Radford S.\,J.\,E., Jewell~P.\,R., Mangum
J.\,G., Bally J.\ eds, 25 years of millimeter spectroscopy. Proc. IAU 170, Kluwer,
Dordrecht, p.\,221

\bibitem[\protect\citename{blah}%
]{WIRE}
Shupe D.\,L. et al., 1998, in McLean B.\,J., Golombek D.\,A., Hayes J.\,J.\,E., 
Payne H.\,E. eds, New horizons with multi wavelength sky surveys. 
Proc. IAU 179, Kluwer, Dordrecht, p.\,118 


\bibitem[\protect\citename{blah}%
]{SIB}
Smail I., Ivison R.\,J., Blain A.\,W., 1997, ApJ, 490, L5 (SIB)

\bibitem[\protect\citename{blah}%
]{SIBK}
Smail I., Ivison R.\,J., Blain A.\,W., Kneib J.-P., 1998, ApJ, 507, L21
(astro-ph/9806061) 

\bibitem[\protect\citename{blah}%
]{S98}
Smail I. et al, 1999, in preparation

\bibitem[\protect\citename{blah}%
]{SP}
Stark A.\,A., Carlstrom J.\,E., Israel F.\,P., Menten K.\,M., Peterson~J.\,B., Phillips
T.\,G., Sironi G., Walker W.\,W., 1998, in Phillips T.\,G.\ ed., Advanced Technology
MMW, Radio and Terahertz telescopes. Proc. SPIE vol. 3357, SPIE, Bellingham,
p. 495 (astro-ph/9802326)

\bibitem[\protect\citename{bl}%
]{StI}
Steidel C.\,C., Giavalisco M., Dickinson M., Adelberger K.\,L., 1996a, AJ, 112, 352

\bibitem[\protect\citename{bl}%
]{StII}
Steidel C.\,C., Giavalisco M., Dickinson M., Pettini M., Dickinson~M., 
Adelberger K.\,L., 1996b, ApJ, 462, L17

\bibitem[\protect\citename{xx}%
]{StL}
Storrie-Lombardi L.\,J., McMahon R.\,G., Irwin M.\,J., 1996, MNRAS, 283, L79

\bibitem[\protect\citename{xx}%
]{TT}
Thronson H., Telesco C., 1986, ApJ, 311, 98

\bibitem[\protect\citename{xx}%
]{TM}
Tresse L., Maddox S.\,J., 1998, ApJ, 495, 691

\bibitem[\protect\citename{xx}%
]{T}
Treyer M.\,A., Ellis R.\,S., Milliard B., Donas J., Bridges T.\,J., 1998, MNRAS, 
300, 303 



\bibitem[\protect\citename{xx}%
]{Web}
Webster R.\,L., Francis P.\,J., Peterson B.\,A., Drinkwater M.\,J., Masci F.\,J., 1995,
Nat, 375, 469

\bibitem[\protect\citename{bl}%
]{WW}
Wilner D.\,J., Wright M.\,C.\,H., 1997, ApJ, 488, L67

\bibitem[\protect\citename{bl}%
]{W}
Wright E.\,L. et al.\ 1994, ApJ, 420, 450

\bibitem[\protect\citename{bl}%
]{X}
Xu C. et al., 1998, ApJ, in press (astro-ph/9806194)

\bibitem[\protect\citename{bl}%
]{ZS}
Zepf S.\,E., Silk J.\,E., 1996, ApJ, 466, 114

\end{thebibliography}
\end{document}